\documentclass[final,5p,times,twocolumn,authoryear]{elsarticle}
\usepackage{amssymb}
\usepackage{lipsum}

\journal{Astronomy $\&$ Computing}

\usepackage{graphicx}
\usepackage{amsmath}
\usepackage{listings}
\usepackage[frozencache,cachedir=.]{minted}
\usepackage{minted}
\usepackage{multicol}
\usepackage{hyperref}

\usepackage[T1]{fontenc}

\usepackage{graphicx}	
\usepackage{amsmath}	
\usepackage{pgfplots}
\usepackage{tikz}
\usepackage{xcolor}

\usetikzlibrary{matrix} 

\pgfplotsset{compat=newest}   

\newcommand{\apachearrow}{\mbox{\textsc{Apache Arrow}}}
\newcommand{\arcae}{\mbox{\textsc{Arcae}}}
\newcommand{\arrow}{\mbox{\textsc{Arrow}}}
\newcommand{\astropy}{\mbox{\textsc{AstroPy}}}

\newcommand{\dask}{\mbox{\textsc{Dask}}}
\newcommand{\daskdistributed}{\mbox{\textsc{Dask Distributed}}}
\newcommand{\pydata}{\mbox{\textsc{PyData}}}
\newcommand{\daskms}{\mbox{\textsc{Dask-MS}}}
\newcommand{\datashader}{\mbox{\textsc{DataShader}}}
\newcommand{\codexafricanus}{\mbox{\textsc{codex africanus}}}
\newcommand{\cupy}{\mbox{\textsc{cuDF}}}
\newcommand{\cudf}{\mbox{\textsc{CuPy}}}
\newcommand{\katdal}{\mbox{\textsc{katdal}}}
\newcommand{\numpy}{\mbox{\textsc{NumPy}}}
\newcommand{\numba}{\mbox{\textsc{Numba}}}
\newcommand{\pandas}{\mbox{\textsc{Pandas}}}
\newcommand{\scipy}{\mbox{\textsc{SciPy}}}
\newcommand{\tensorflow}{\mbox{\textsc{TensorFlow}}}

\newcommand{\jax}{\mbox{\textsc{Jax}}}
\newcommand{\numfocus}{\mbox{\textsc{NumFOCUS}}}
\newcommand{\openmp}{\mbox{\textsc{OpenMP}}}
\newcommand{\plotms}{\mbox{\textsc{plotms}}}
\newcommand{\pytorch}{\mbox{\textsc{PyTorch}}}
\newcommand{\casa}{\mbox{\textsc{CASA}}}
\newcommand{\xarray}{\mbox{\textsc{XArray}}}
\newcommand{\zarr}{\mbox{\textsc{Zarr}}}
\newcommand{\cubical}{\mbox{\textsc{CubiCal}}}
\newcommand{\QuartiCal}{\mbox{\textsc{QuartiCal}}}
\newcommand{\pfbimager}{\mbox{\textsc{pfb-imaging}}}
\newcommand{\ddfacet}{\mbox{\textsc{DDFacet}}}
\newcommand{\wsclean}{\mbox{\textsc{WSClean}}}
\newcommand{\killms}{\mbox{\textsc{killMS}}}
\newcommand{\meqtrees}{\mbox{\textsc{MeqTrees}}}
\newcommand{\oldstimela}{\mbox{\textsc{Stimela}}}
\newcommand{\tricolour}{\mbox{\textsc{Tricolour}}}
\newcommand{\shadems}{\mbox{\textsc{ShadeMS}}}
\newcommand{\stimela}{\mbox{\textsc{Stimela2}}}
\newcommand{\ragavi}{\mbox{\textsc{RAGaVI}}}

\DeclareRobustCommand{\VAN}[3]{#2}
\let\VANthebibliography\thebibliography
\def\thebibliography{\DeclareRobustCommand{\VAN}[3]{##3}\VANthebibliography}

\begin{document}
\begin{frontmatter}



\title{Africanus I. Scalable, distributed and efficient radio data processing with Dask-MS and Codex Africanus}
\begin{abstract}
The physical configuration of new radio interferometers such as MeerKAT, SKA, ngVLA and DSA-2000 informs the development of software in two important areas. Firstly, tractably processing the sheer quantity of data produced by new instruments necessitates subdivision and processing on multiple nodes. Secondly, the sensitivity inherent in modern instruments due to improved engineering practices and greater data quantities necessitates the development of new techniques to capitalise on the enhanced sensitivity of modern interferometers.

This produces a critical tension in radio astronomy software development: a fully optimised pipeline is desirable for producing science products in a tractable amount of time, but the design requirements for such a pipeline are unlikely to be understood upfront in the context of artefacts unveiled by greater instrument sensitivity. Therefore, new techniques must continuously be developed to address these artefacts and integrated into a full pipeline. As Knuth reminds us, ``Premature optimisation is the root of all evil''. This necessitates a fundamental trade-off between a trifecta of (1) performant code (2) flexibility and (3) ease-of-development. At one end of the spectrum, rigid design requirements are unlikely to capture the full scope of the problem, while throw-away research code is unsuitable for production use.

This work proposes a framework for the development of radio astronomy techniques within the above trifecta. In doing so, we favour flexibility and ease-of-development over performance, but this does not necessarily mean that the software developed within this framework is slow. Practically this translates to using data formats and software from the Open Source Community. For example, by using \numpy\ arrays and/or \pandas\ dataframes, a plethora of algorithms immediately become available to the scientific developer.

Focusing on performance, the breakdown of Moore's Law in the 2010s and the resultant growth of both multi-core and distributed (including cloud) computing, a fundamental shift in the writing of radio astronomy algorithms and the storage of data is required:
It is necessary to {\it shard} data over multiple processors and compute nodes, and to write algorithms that operate on these shards in parallel.
The growth in data volumes compounds this requirement.
Given the fundamental shift in compute architecture we believe this is central to the performance of any framework going forward, and is given especial emphasis in this one.


 This paper describes two Python libraries, \daskms\ and \codexafricanus\, which enable the development of distributed High-Performance radio astronomy code with \dask. \dask\ is a lightweight Python parallelisation and distribution framework that seamlessly integrates with the  \pydata\ ecosystem to address radio astronomy ``Big Data`` challenges.

\end{abstract}


\author[sarao]{S.~J.~Perkins}\ead{simon.perkins@gmail.com}
\author[ratt]{J.~S.~Kenyon}
\author[ratt]{L.~A.~L. ~Andati}
\author[sarao,ratt]{H.~L.~Bester}
\author[ratt,sarao,ira]{O.~M.~Smirnov}
\author[sarao,ratt]{B.~V.~Hugo}

\affiliation[sarao]{organization={South African radio astronomy Observatory (SARAO)},
            city={Cape Town},
            state={WC},
            country={South Africa}}

            \affiliation[ratt]{organization={Centre for radio astronomy Techniques \& Technologies (RATT),
Department of Physics and Electronics, Rhodes University},
            city={Makhanda},
            state={EC},
            country={South Africa}}

\affiliation[ira]{organization={Institute for Radioastronomy, National Institute of Astrophysics (INAF IRA)},
            city={Bologna},
            country={Italy}}

\end{frontmatter}

\section{Introduction}

The new generation of ``SKA precursor'' radio telescopes such as MeerKAT \citep{Jonas2016}, LOFAR \citep{van_Haarlem_2013} and ASKAP \citep{Hotan_2021}, as well as the ongoing development of SKA Phase 1 \citep{schilizzi2008square}, ngVLA \citep{ngvla2019} and DSA-2000 \citep{dsa2000} has spurred active research into new radio astronomy data processing algorithms and techniques which can take full advantage of the revolutionary capabilities of such instruments. At the same time, the sheer volume of data produced by these observatories requires highly parallel and distributed processing. This has produced a critical tension in radio astronomy software development. On the one hand, fully optimised, distributed software pipeline implementations are desirable for producing science products in a tractable amount of time. On the other hand, the design requirements and underlying algorithms for such pipelines represent a continuously moving target. This has created a landscape where ``research software'', developed rapidly by relatively small teams, needs to be performant enough to cope with full-scale precursor data in order for new algorithms to be fully validated. Conversely, sufficiently performant ``research software'' can sometimes see uptake into operational pipelines, in preference to a fully optimized reimplementation that may be obsoleted by newer ``research software'' before it is complete. It is telling, for example, that the LOFAR MSSS pipeline \citep{msss2018} is built around the \ddfacet/\killms\ \citep{ddfacet2018} packages -- a classic case of the uptake of ``research software'' that was not even yet conceptualized algorithmically at the time of LOFAR's official inauguration in 2010.

{\bf The breakdown of Moore's law and the growth of multi-processing and distributed computing:} In the 2010s, due to a combination of physics and economics, it became difficult for chip manufacturers to continue extracting performance gains by increasing both transistors counts and clock speeds of individual processors \citep{Alted2010}.
To remain competitive, manufacturers began placing multiple processors on a single die, leading to the current era of multi-core processors, most notably GPU's whose architecture is predicated on many cores executing Single Instructions on Multiple Data (SIMD). Thus, to efficiently use a modern CPU, algorithms must be {\it sharded} over its processors, and distributed computing extends this concept further by {\it sharding} algorithms over multiple nodes in a compute cluster.
Sharding, therefore, is critical to the development of modern, performant, distributed algorithms and file systems.
Unfortunately, both AIPS \citep{aips1990} and CASA \citep{McMullin2007} were developed well-before or just on the cusp of this change in architecture.
For this reason, we argue that a new generation of radio astronomy software is required that takes these changes in computer architecture and distributed computing paradigms into account from the outset.
Fortunately, we believe that it is still possible to incorporate many of the excellent design decisions made by these earlier ecosystems.

{\bf The Python ecosystem:} ``Research software'' necessitates a fundamental trade-off between a trifecta of (1) reasonably performant code, (2) flexibility and (3) ease-of-development. The Python language, and the surrounding software ecosystem, has filled a similar niche in the Big Data/ML space. The so-called \pydata\ community (which, we should note, is much bigger than the radio astronomy software community) is driving very rapid development of novel Python-based software technologies, which has allowed for HPC-scale implementations of tools with performance comparable to  that of traditional (e.g. C/C++ MPI) handcrafted HPC code, while retaining the development agility of a high-level interpreted language.

Python has also seen widespread adoption in radio astronomy.
Initially, it filled the role of a high-level scripting language in packages
such as \meqtrees\ \citep{Noordam2012}, \casa\ \citep{McMullin2007}
and Parseltongue \citep{Kettenis2006}.
The emergence of high-performance numerical libraries (\numpy/\scipy)
and the \astropy\ Project \citep{Astropy2022} drove the use of Python in
radio data \emph{analysis}, particularly in the space of data products,
which tend to be smaller in volume than the raw data that \emph{reduction}
software deals with.
The {\tt multiprocessing} and {\tt concurrent.futures} modules of the
Python Standard Library (PSL) allowed it to exploit multi-core parallelism efficiently.
This has ultimately enabled the development of fully (or nearly fully)
Pythonic calibration and imaging packages such as \cubical\ \citep{kenyon2018cubical}
and the above-mentioned \ddfacet/\killms\ suite, which lie squarely in the data reduction space,
and provide sufficient performance to successfully deal with raw precursor data.
However, up to now radio astronomy has been slow to leverage the software technologies
coming out of the \pydata\ community, and to take advantage of the technical innovation therein.
This series of papers, and the software packages underpinning them, aims to change that.

In particular, this approach is inspired by the {\bf Pangeo} project \citep{Abernathey2017}
which pioneered the processing of large quantities of
climate data using the \pydata{} ecosystem on the cloud.
Pangeo climate data is stored as \zarr\ \citep{Zarr2018} arrays on cloud object stores
such as Google Cloud Platform and Amazon S3.
This data is exposed to the developers as \xarray\ \citep{xarray2017} datasets of
\dask\ \citep{Dask2016} arrays which are then distributed and processed in
parallel across multiple cloud instances.

{\bf Formats and Interfaces:}
One of the core features contributing to the success of \numpy, \scipy\ and \pandas\
is the creation of standardised formats and interfaces on which a wide range of algorithms
can operate, namely the multi-dimensional \numpy\ and \cupy\ arrays as well as the \pandas\ dataframe.
As they are well-defined, their contents are easily exposed to other programming languages,
most notably C/C++ extensions.
Easily accessible formats and interfaces make it possible to
quickly link algorithms together.
They also enable for conversion between other
formats, enabling rapid development crucial to testing out ideas and algorithms.

Interestingly, the formats central to the radio data ecosystem,
FITS \citep{Wells1981} and the CASA Table Data System (CTDS) \citep{Diepen2015},
roughly correspond to \numpy\ arrays and \pandas\ dataframes, respectively.
The CTDS-based Measurement Set exhaustively specifies the format
for observational data, especially {\emph visibilities}, the complex-valued correlated power between pairs of radio receivers.
However, both FITS and CTDS were implemented prior to the advent of
modern distributed and cloud computing, and do not cleanly implement
distributed file locking, or multi-threaded access.
Fortunately, both the FITS and CTDS specifications map easily onto contemporary
cloud-native formats such as \zarr \citep{Zarr2018}
and \apachearrow \citep{Arrow2019}.
These formats offer support for multiple, multi-threaded readers and writers
which is largely achieved by subdividing data into separate files, thereby avoiding
lock contention on single, monolithic files.
Additionally, the aforementioned formats and implementations are used,
supported and maintained by multiple, well-resourced organisations.

{\bf Containerization:} Another pertinent development in computing is
the emergence of container technology, beginning with the widespread adoption of
\textsc{Docker}\footnote{\url{https://docker.com}}, with other container engines (\textsc{Singularity}, \textsc{Podman})
also gaining popularity. A container is essentially a ``virtual machine lite'', wrapping and
isolating (for example) an application complete with its dependencies, down to the
operating system level. Container {\em images} can be deployed and instantiated across a variety
of computing environments.
This provides the ability to ship and execute software in a platform-independent way,
largely taking away the ``dependency hell'' and version conflict problem that used to
arise when attempting to install a diverse set of software packages on a single platform.
Containerization is particularly well-suited to research software, which, by its nature,
often tends to rely on a tangled web of dependency libraries with specific versioning requirements.
Containers are gaining popularity in radio astronomy, with efforts such as
\oldstimela\footnote{\url{https://github.com/ratt-ru/Stimela}} \citep{makhathini2018} and
the \textsc{KERN} suite\footnote{\url{https://kernsuite.info}} \citep{Molenaar2021} providing
sets of container images for popular radio astronomy packages. Paper IV in this series \citep{africanus4}
presents \stimela, a new-generation workflow management framework to which containerization is central.

{\bf Cloud computing:} Sometimes referred to as {\em commodity compute}, cloud computing supports the provision of a generic set of computing services over the internet.
Commercial cloud providers such as Amazon, Google and Microsoft 
host large-scale data centers that provide on-demand computing and storage in the
form of virtual machines (e.g. Amazon EC2), storage endpoints (e.g. Amazon S3),
and various application-specific services such as database servers.
Academic HPC centres are also experimenting with providing some of
their resources as cloud-type services.
Basic economy-of-scale arguments suggest that, in an ideal world,
the unit cost of compute and storage from a huge vendor such as Amazon and Google
must be cheaper than deploying on-premises hardware for all but the largest organizations.
Practicalities such as data transfer and the fact that even idle cloud resources
incur billing mean that these cost advantages can be non-trivial to realize,
particularly in the research software space.
However, the Rubin Observatory (LSST), a 100 PB-scale project,
has deployed its Science Data Platform on Google Cloud \citep{rubin2023}.
In the radio astronomy world, several efforts
\citep{cloudpulsar2017,cloudlofar2017,cloud21cmcosmology2021,dodson-cloud,dodson-cloud2}
have demonstrated use of the cloud for some project-specific data reduction.
However, efficiently deploying generic radio astronomy pipelines in the cloud
remains an elusive goal.

A notable development at the intersection of containerization and cloud computing is the emergence of container cluster technology such as Kubernetes (a.k.a K8s). A K8s cluster provides compute resources in the form of container {\em pods} that the K8s scheduler maps to available compute infrastructure (which can be anything from a single compute node, to an on-premises cluster, to virtual machine instances in the cloud). Later in the series we show how our work leverages K8s to take an important step towards the above-mentioned elusive goal.

{\bf About the series:} Paper I presents the \daskms\footnote{\url{https://github.com/ratt-ru/dask-ms}} and \codexafricanus\footnote{\url{https://github.com/ratt-ru/codex-africanus}} packages, which provide both the data access layer and the fundamental building blocks for a new generation of highly parallel, scalable and distributed data reduction and analysis packages that can run on a range of architectures, from an end-user's laptop, to ``bare metal'' HPC nodes, to cloud infrastructure such as AWS. Some of these smaller packages are covered in this paper. Two of the larger (and algorithmically novel) applications, \QuartiCal\footnote{\url{https://github.com/ratt-ru/QuartiCal}} (calibration suite) and \pfbimager\footnote{\url{https://github.com/ratt-ru/pfb-clean}} (imager), are described in separate Papers II and III. Collectively, this software suite provides all the components for a complete calibration and imaging pipeline. Paper IV deals with the \stimela\ workflow management framework that supports the implementation of such pipelines, and allows their deployment either ``natively'' (i.e. on a single local compute node), across a Slurm cluster, or on a K8s cluster, which can be either local, or hosted in the cloud. Upcoming work will focus on the deployment of such a pipeline on the AWS cloud, and describe efficient solutions for remote visualization of data products and other pipeline outputs.

\section{Background}
\label{section:background}

{\bf AIPS and CASA:} Broadly speaking, two generations of radio astronomy software exist, The Astronomical Image Processing System (AIPS) \citep{aips1990} was developed in the late 1970's, and Common Astronomy Software Applications (CASA) \citep{McMullin2007} in the late 1990's.
Other suites exist and are extensively documented in \citep{Molenaar2021}, but the above two are important because they are the most widely used and also because they both established commonly used formats: the Flexible Image Transport System (FITS) \citep{Wells1981} and the CASA Table Data System (CTDS) \citep{Diepen2015}\footnote{Strictly speaking, AIPS did not establish the FITS standard per se, but rather established the adoption of this standard in radio astronomy. CASA did establish CTDS from the ground up, via its precursor package AIPS++.}.
These formats have played the crucial role of a \emph{lingua franca} in the world of radio astronomy by allowing different software packages to communicate with each other.
The FITS format describes a series of multi-dimensional binary arrays and their associated metadata, and can be used to store both the visibility and image data on-disk. FITS Extensions allow the storage of related data such as Spectral Window information in sub-tables. The simplicity of FITS is a great strength, and the \numpy\ array we describe later is a similarly simple construct. For this reason, it is loved by astronomers. 
Due to the amount of data produced by newer instruments, FITS files became inappropriate for storing large quantities of visibilities \citep{Molenaar2021} and work began on defining {\it Measurement Sets} based on the CTDS \citep{Diepen2015}. The Measurement Set specification \citep{MeasurementSet2}  exhaustively describes observational data produced by many different types of radio interferometers.
However, the CTDS was implemented before the era of multi-core, distributed computing and has characteristics that do not fit well with this paradigm:

\begin{enumerate}
    \item Multi-threaded access to the CTDS was never implemented.
    \item Columns are stored in monolithic files which, strictly speaking, should be locked when accessed by multiple readers/writers. However, in distributed settings, this is worked around by disabling locking and requiring software to not create data races, {\it by convention}.
    \item The CTDS does not support ACID (atomicity, consistency, isolation, durability) properties. It is possible to inadvertently corrupt CASA Tables during application failure, losing all progress and requiring restarting reductions from scratch.
    \item It has a POSIX implementation which is incompatible with the cheaper object stores of cloud-native storage systems like AWS S3.
\end{enumerate}

{\bf Strong and Weak scaling:}
Radio astronomy is fundamentally a big data problem and as such, it is amenable to horizontal scaling. Amdahl's Law \citep{Amdahl1967} and Gustafson's \cite{Gustafson1990} Law are two fundamental principles used to analyze and predict the performance of parallel computing systems in the context of strong and weak scaling. Both laws offer insights into how much speedup can be achieved by parallelizing a computational problem and how well the system can handle an increasing workload.

Amdahl's Law focuses on improving the performance of a fixed problem size by parallelizing a portion of the computation. The law provides a theoretical speedup limit for parallel processing. It is expressed as:
\begin{equation}
\textrm{SPEEDUP}(N) = \frac{1}{(1-P) + \frac{P}{N}}
\end{equation}

Where $\textrm{SPEEDUP}(N)$ is the improvement in execution time achieved by parallelizing the computation.
$P$ is the fraction of the program that can be parallelized (ranging from 0 to 1) and $N$ is the number of processors (computational units) used for parallel execution.

The speedup is limited by the sequential part of the computation (1 - P). As the number of processors increases, the impact of the non-parallelizable portion becomes more pronounced.
To achieve significant speedup, the parallelizable portion of the problem (P) must be maximized.

Instead of keeping the problem size fixed, Gustafson's Law focuses on scaling the workload proportionally to the number of processors. It addresses the idea that larger problems may contain more parallelizable portions and that parallel systems should handle increasing workloads efficiently. The law is expressed as:
\begin{equation}
\textrm{SPEEDUP}(N) = N + S \times (1 - N)
\end{equation}

Where $\textrm{SPEEDUP}(N)$ is the improvement in execution time achieved by parallelizing the computation, $N$ is the number of processors and $S$ is the fraction of time spent on the serial portion of the programme.

In contrast to Amdahl's Law, Gustafson's Law suggests that larger problems tend to have a higher percentage of parallelizable work, which can lead to better scaling with more processors.
As the number of processors increases, the sequential part of the problem becomes less significant, and the system can efficiently handle larger workloads.
In summary, Amdahl's Law emphasizes the importance of optimizing the parallelizable portion of a problem to achieve speedup, while Gustafson's Law highlights the potential benefits of scaling the workload proportionally with the number of processors to handle larger problems efficiently. Both laws provide valuable insights into the performance characteristics of parallel computing systems and help guide the design and optimization of parallel algorithms and architectures.

{\bf The \pydata\ ecosystem:}
Python's ease-of-development has contributed to its popularity in the scientific community.
In particular, \numfocus, a non-profit organisation promoting open practices in research and scientific computing, has been established through corporate sponsorship from entities such as Google and NVIDIA. \numfocus\ has sponsored the continued development of many open-source libraries,
including \numpy\ \citep{Harris2020}, \scipy\ \citep{Virtanen2020},
\pandas\ \citep{pandas2010}, and, in the case of astronomy, the convenient \astropy\ \citep{Astropy2022}.
These libraries make it easy for scientific developers to rapidly write code,
as well as publish it and make it available to other developers.

The \numpy\ multi-dimensional array (ndarray), and the \pandas\ dataframe
are core data structures on which a wide variety of algorithms are built.
These data structures and algorithms are implemented as C/C++ extensions,
but are exposed as Python objects, allowing scientific programmers to easily manipulate them.
\numpy\ arrays are, simply, memory buffers with associated dimension metadata,
while \pandas\ dataframes collate several ndarrays that share a primary ``row'' index, or coordinate.
This has resulted in large organisations writing extensive Open Source frameworks for use by the wider community.
Examples of this include \tensorflow\ and \pytorch\,
written by Google and Meta, respectively.
Even though much of the performant code is written in C++, familiar Python
interfaces are provided which ingest and consume \numpy\ arrays.

Python's ease-of-development and flexibility involves tradeoffs.
It is a dynamically typed, interpreted language whose standard CPython
interpreter only allows a single thread to interpret Python byte code
at a time.
Interpreted languages are several orders of magnitude slower than
languages compiled directly to machine code.
Also, to interpret Python byte code, threads must acquire the global interpreter lock (GIL),
severely limiting Python's multi-threading capabilities.
Attempts to remove or work-around the GIL have proliferated within the Python community. Recent attempts include PEP 554\footnote{\url{https://peps.python.org/pep-0554}}, and the ``nogil'' project\footnote{\url{https://github.com/colesbury/nogil}} started by a \pytorch\ developer.
Applications can also use multiple processes, each running their own Python interpreter, but there are either (1) serialisation overheads for inter-process communication or (2) implementation and configuration nuances inherent in the use of POSIX shared memory systems.

{\bf \numba: } Other strategies for avoiding the GIL involve writing C/C++ extensions, but these require scientists to learn and write code in complex, unforgiving languages.
An exciting compromise is the \numba\ \citep{Lam2015} project, which just-in-time (JIT) compiles functions written in a subset of Python and \numpy\  into machine code with performance comparable to C/C++.
It is also possible to write NVIDIA CUDA kernels in \numba\ to exploit high performance GPU targets.
This performance is obtained because \numba\ (along with the Rust and Clang C++ compilers) uses the LLVM compiler framework to emit machine code.
When a \numba\ decorated Python function is called, it is dispatched to an implementation unique to the supplied argument types and JIT compiled.
This {\it multiple dispatch} mechanism also works in conjunction with simple literal arguments (integers, floats and strings).
Given that JIT'd functions can call other JIT'd functions, multiple dispatch provides an extraordinary degree of flexibility
when composing performant computation kernels from individual components.
Julia \citep{Julia-2017} has this capability built in at the language level,
but similar flexibility and performance would likely only be achievable in C++
via the use of template metaprogramming (and the associated compiled template bloat).

JIT'd functions are cacheable by argument types to prevent needless compilation on multiple calls, and cached functions persist between application runs.
Additionally, since functions are JIT'd, there is no need to package binary wheels for a wide range of computer architectures, easing the burden
of package maintenance.
\numba\ can of course compile functions that ingest and use \numpy\ arrays, as well as most Python primitive types.
Finally, because it executes machine code, \numba\ functions can drop the GIL, allowing other threads to interpret Python byte code (or run other machine code) while it executes.
This works well in a radio astronomy data reduction environment where \numba\ functions can spend several seconds processing a portion of the data.

{\textsc{Jax}} \citep{jax2018} is a Machine-Learning framework
that uses the XLA compiler \citep{OpenXLA} to JIT compile \numpy\ code
to CPU, GPU and Tensor Processing Unit (TPU) architectures.
It also supports automatic differentiation (autodiff), auto-vectorisation
and parallel programming.
As such it provides a convenient mechanism for writing code that
can easily take advantage of accelerators.
One idiosyncrocy of XLA is that it JIT-compiles a function for
every different shape of the input tensor arguments as this allows
it to optimally tile data so that it can fit optimally into processor caches.
For data whose shape varies, this can result in spending more time JIT-compiling,
or padding data.
Jax arrays are immutable, so algorithms that modify data in-place
are not supported.
This means that functions written in Jax can use more memory
than those written in C++ or \numba, for example.
Additionally, it is a heavy dependency, requiring ~500MB of disk
space for a Python 3.10 x86\_64 install on Ubuntu 22.04.

{\bf Dataflow programming.} Radio astronomy is inherently a Big Data problem, requiring tools to conveniently and efficiently exploit multiple processors and cluster nodes. Dataflow programming \citep{Dennis1974a, Dennis1974b, Rodrigues1969} is a programming paradigm where programs are modelled as graphs through which data flows. Each node represents a task that (a) ingests data from other nodes linked by a graph edge (b) transforms it and (c) outputs data for consumption by other nodes. Compilers use these structures to represent program structure and flow, and thereby assign variables to scarce but fast registers.

\meqtrees\ \citep{Noordam2012} pioneered dataflow programming within radio astronomy, but the style has seen a general resurgence during the 2010s. This can be attributed to a couple of factors. Firstly, new machine learning frameworks such as \tensorflow\ and \pytorch\ need to  minimise loss functions and it is efficient to evaluate function gradients via back-propagation on dataflow graphs. Secondly, it provides a safe, convenient and functional style for writing multi-process programs that are now required to fully exploit contemporary multi-core processors. It is difficult, even for experienced developers, to write correct multi-process code free from race-conditions and deadlocks in an imperative style.
By contrast, representing a program as a graph allows task dependencies to be inferred. Consequently, tasks can be independently scheduled on multiple processors, eliminating the need for writing complex multi-process code.
This necessarily delegates responsibility for program control flow and task placement from the developer to a scheduler.

{\bf \dask\ } \citep{Dask2016} is a Python \pydata\ dataflow library for parallel and distributed computing. Low-level \dask\ graphs are represented as Python dictionaries where each key is a unique node identifier as shown in Listing \ref{listing:dict_graph} The associated value can be either a literal or a task where the latter is defined by a tuple with a callable function as the first element. Subsequent elements contain arguments represented as literals, or the node identifiers of dependencies.


\newlength{\originalcolumnsep}
\setlength{\originalcolumnsep}{\columnsep}
\setlength{\columnsep}{-2.5cm}
\begin{listing}
\begin{multicols}{2}
\begin{minted}{python}
a = 1
b = 2
c = a + b
d = b + 1
e = c * d
assert e == 9
\end{minted}
\begin{minted}{python}
dsk = {'a': 1,
       'b': 2,
       'c': (sum, ['a', 'b']),
       'd': (sum, ['b', 1]),
       'e': (mul, 'c', 'd')}
assert dask.get(dsk, "e") == 9
\end{minted}
\end{multicols}
\caption{A piece of code written imperatively, and as a low-level \dask\ graph.}
\label{listing:dict_graph}
\end{listing}

\begin{listing}
\begin{minted}{python}
dsk = {
    ("A-a53d", 0, 0): (np.ones, (100, 50)),
    ("A-a53d", 0, 1): (np.ones, (100, 50)),
    ("A-a53d", 1, 0): (np.ones, (100, 50)),
    ("A-a53d", 1, 1): (np.ones, (100, 50))}
chunks = ((100, 100), (50, 50))  # per dimension
A = dask.Array("A-a53d", dsk, chunks, np.float64)
assert A.shape == tuple(map(sum, chunks))
\end{minted}
\caption{High level \dask\ Array collections are formed
from a low-level graph, and associated metadata in the
form of a unique name, per-dimension chunks and a data type.
In this example, keys have the form $(\textrm{name}, i, j)$
where name is name is a unique name identifying the
array collection, while $i$ and $j$ are chunk identifiers for
the first and second dimension.
Each key is associated with a task describing the parameters
for calling {\tt np.ones()}.}
\label{listing:collections}
\end{listing}

\begin{listing}
\begin{minted}{python}
import dask.array as da

# Lazily construct million element dask arrays
# partitioned into 100 chunks
A = da.ones((1000, 1000),
             chunks=(100, 100),
             dtype=np.float64)
B = da.random.random((1000, 1000),
                     chunks=(100, 100))
# Lazily construct expression C
C = (A * B).sum(axis=1)
# Evaluate expression C
print(C.compute())
\end{minted}
\caption{In the above code snippet, two \dask\ arrays are created
and multiplied together before a summation reduction is
performed on the last axis.
Each chunk is associated with a graph node.}
\label{listing:dask_arrays}
\end{listing}

\setlength{\columnsep}\originalcolumnsep

The simplicity of this definition supports strong integration
with other \pydata\ ecosystem packages, especially \numpy\ and \pandas.
However, instead of requiring a developer to manually construct a graph,
\dask\ provides useful {\it Collections} that simplify this process.
Most notably, \dask\ arrays and dataframes
which have similar semantics to their \numpy\ and \pandas\ counterparts.
These collections present the scientific programmer with a familiar interface for
developing parallel and distributed applications, as shown in Listing~\ref{listing:dask_arrays}.

Instead of strictly representing data, these collections encapsulate lightweight metadata and an associated computational graph, thereby describing {\it a series of operations for generating data.}
Encoded in this metadata are \textit{chunks}: a list of heterogeneous sizes
that subdivide each array dimension. Each chunk is associated with a
graph node and this provides the primary mechanism
for expressing \textit{data parallelism} within \dask\, as shown in Listing~\ref{listing:collections}.

\dask\ Arrays are also \textit{lazily-evaluated}: The scientific programmer can create complex expressions involving \dask\ arrays but no computation occurs until explicitly requested. Graph nodes unrelated to the final result are also not executed.
Therefore, \dask\ array expressions are created with \numpy\ semantics, but in practice create a graph, or program that is not executed until the result is requested. This means that \dask\ arrays are both \textit{programs} and \textit{interfaces} to other programs, supporting flexible composition of arbitrary operations.

\dask\ graphs can be scheduled using multi-threaded, multi-process and
distributed schedulers.
The first two target Python threads
and processes, depending on the degree to which the executed functions hold the GIL.
The distributed scheduler is more sophisticated and
schedules tasks on multiple cluster nodes, enabling \textit{horizontal scaling} of applications.

In addition to the appealing Collections interface, \daskdistributed\ \footnote{\url{https://distributed.dask.org}} supports a {\it Distributed Task Executor}.
Developers create {\it Client} objects, to which a function (task) and associated arguments can be submitted for execution on a distributed cluster of workers.
A {\it Distributed Future}, representing the result of said computation is instantaneously returned and can serve as an argument to other distributed function calls, allowing developers to iteratively compose a distributed graph.
It is also possible to assign tasks to specific workers or constrain tasks to workers
configured with specific resources such as GPUs.
This provides low-level control, with a similar level of complexity
to that of imperatively written MPI \citep{mpi41} code.

Finally, \dask\ provides limited support for {\it actors} \citep{Hewitt1973} which are stateful
distributed objects that can communicate with other distributed entities via messages.

{\bf \xarray:} The \pandas\ dataframe is a tabular data structure that collates one-dimensional \numpy\ arrays related by a single ``row'' index, or coordinate. This representation is highly convenient for data analytics. The \xarray\ project \citep{xarray2017} extends this concept by collating multi-dimensional \numpy\ or \dask\ arrays into an \xarray\ Dataset.
\xarray\ allows for each dimension on a DataArray object to be labelled.
This provides a convenient mechanism for aligning arrays which share dimensions.

\xarray\ Datasets are a convenient abstraction for representing radio astronomy data stored in the
Measurement Set format, as all data is aligned along a primary row dimension,
but visibility data also has secondary channel and correlation dimensions.
The channel dimension could be assigned frequency values as coordinates,
while the correlation dimension could be assigned ``XX'', ``XY'', ``YX'' and ``YY''
labels.
They are also self-describing \citep{Tilmes2011}: information describing the contents
of the Dataset is included in the Dataset itself. This can be seen in
Listing~\ref{listing:xarray}, where Measurement Set data is
represented as \dask\ arrays aggregated into an \xarray\ Dataset.
Additional metadata such as dimension sizes, coordinates, chunk sizes, data types
and attributes describe this data.

\begin{listing*}
\begin{minted}{python}
 Dimensions:         (row: 75600, chan: 64, corr: 4, uvw: 3)
 Coordinates:
     ROWID           (row) int32 dask.array<chunksize=(10000,), meta=np.ndarray>
 Dimensions without coordinates: row, chan, corr, uvw
 Data variables: (12/22)
     ANTENNA1        (row) int32 dask.array<chunksize=(10000,), meta=np.ndarray>
     ANTENNA2        (row) int32 dask.array<chunksize=(10000,), meta=np.ndarray>
     FEED2           (row) int32 dask.array<chunksize=(10000,), meta=np.ndarray>
     OBSERVATION_ID  (row) int32 dask.array<chunksize=(10000,), meta=np.ndarray>
     MODEL_DATA      (row, chan, corr) complex64 dask.array<chunksize=(10000, 64, 4), meta=np.ndarray>
     SIGMA           (row, corr) float32 dask.array<chunksize=(10000, 4), meta=np.ndarray>
     CORRECTED_DATA  (row, chan, corr) complex64 dask.array<chunksize=(10000, 64, 4), meta=np.ndarray>
     ...              ...
     DATA            (row, chan, corr) complex64 dask.array<chunksize=(10000, 64, 4), meta=np.ndarray>
     INTERVAL        (row) float64 dask.array<chunksize=(10000,), meta=np.ndarray>
     EXPOSURE        (row) float64 dask.array<chunksize=(10000,), meta=np.ndarray>
     TIME            (row) float64 dask.array<chunksize=(10000,), meta=np.ndarray>
     FLAG_ROW        (row) bool dask.array<chunksize=(10000,), meta=np.ndarray>
 Attributes:
     FIELD_ID:             0
     DATA_DESC_ID:         0]
\end{minted}
\caption{An \xarray\ Dataset containing \dask\ arrays representing Measurement Set data. Multiple dimensions are represented, most notably row, channel and correlation (row, chan, corr). Separate arrays are aligned along these coordinates.}
\label{listing:xarray}
\end{listing*}

{\bf Distributed cloud-native formats:} The \pydata\ ecosystem has access to
highly performant file formats and implementations.
Amongst other endeavours, the \apachearrow\ project \citep{Arrow2019} defines an open standard for columnar, in-memory and on-disk data that is tightly integrated with \numpy\ arrays and \pandas\ dataframes,
partly due to the fact that the founder of \apachearrow\ also originally created the \pandas\ project.
This standard is implemented  in multiple languages, including C++, Rust and Python, freely allowing zero-copy interchange of data between them.
\arrow\ stores columnar data in a hierarchical directory structure containing avro, parquet or orc files.
\arrow\ column types can be flexibly composed from a combination of primitive (int, floats, strings) and compound types (dictionaries and structures).
While only one-dimensional \numpy\ arrays are supported out of the box, it possible to construct high dimensional tensors and other, more complex types through the use of Extension Types.
Another important factor in favour of \arrow\ is the fact that it is a large ecosystem in its own right and experiencing explosive growth.


\zarr\ \citep{Zarr2018} is an in-memory and on-disk representation for groups of multi-dimensional arrays.
These arrays follow the \numpy\ array interface, allowing users to easily manipulate data both in-memory and on disk.
It has a chunking schema similar to that of \dask\ where each chunk is mapped to a file on disk and,
it is possible to chunk data along multiple dimensions.
A hierarchical directory structure contains directories for both groups and arrays.
Individual array chunks are stored in the array directories.

While the above formats can store data on POSIX file systems, they work particularly well with newer, {\it object storage systems}, which are designed with massive amounts of data and the attendant cost and scalability concerns in mind.
At the interface level, object stores map keys to large binary objects (BLOBs).
Then, by design, both \arrow\ and \zarr\ partition datasets into multiple logical objects that can be read and written independently by multiple  nodes, processes or threads, reducing locking contention.
As radio astronomy data sizes grow, the cheaper cost, and greater throughput and scalability of object storage will quickly outweigh the advantages of traditional file systems, to which the CTDS is currently limited.
Additionally, object stores offer fine-grained performance control and cost monitoring.
For example, greater S3 throughput can be provisioned for more money,
and archival data can be moved to much cheaper long-term S3 Glacier storage,
at the cost of a longer retrieval time. Notably, the MeerKAT archive\footnote{\url{https://archive.sarao.ac.za}} uses a Ceph object store under the hood, with data converted to CTDS only upon export via the \katdal\ \citep{Schwardt2023} data access layer.


\section{Designing a new radio astronomy software ecosystem}
\label{section:ecosystem}

\begin{figure*}
\begin{center}
\includegraphics[width=0.75\textwidth]{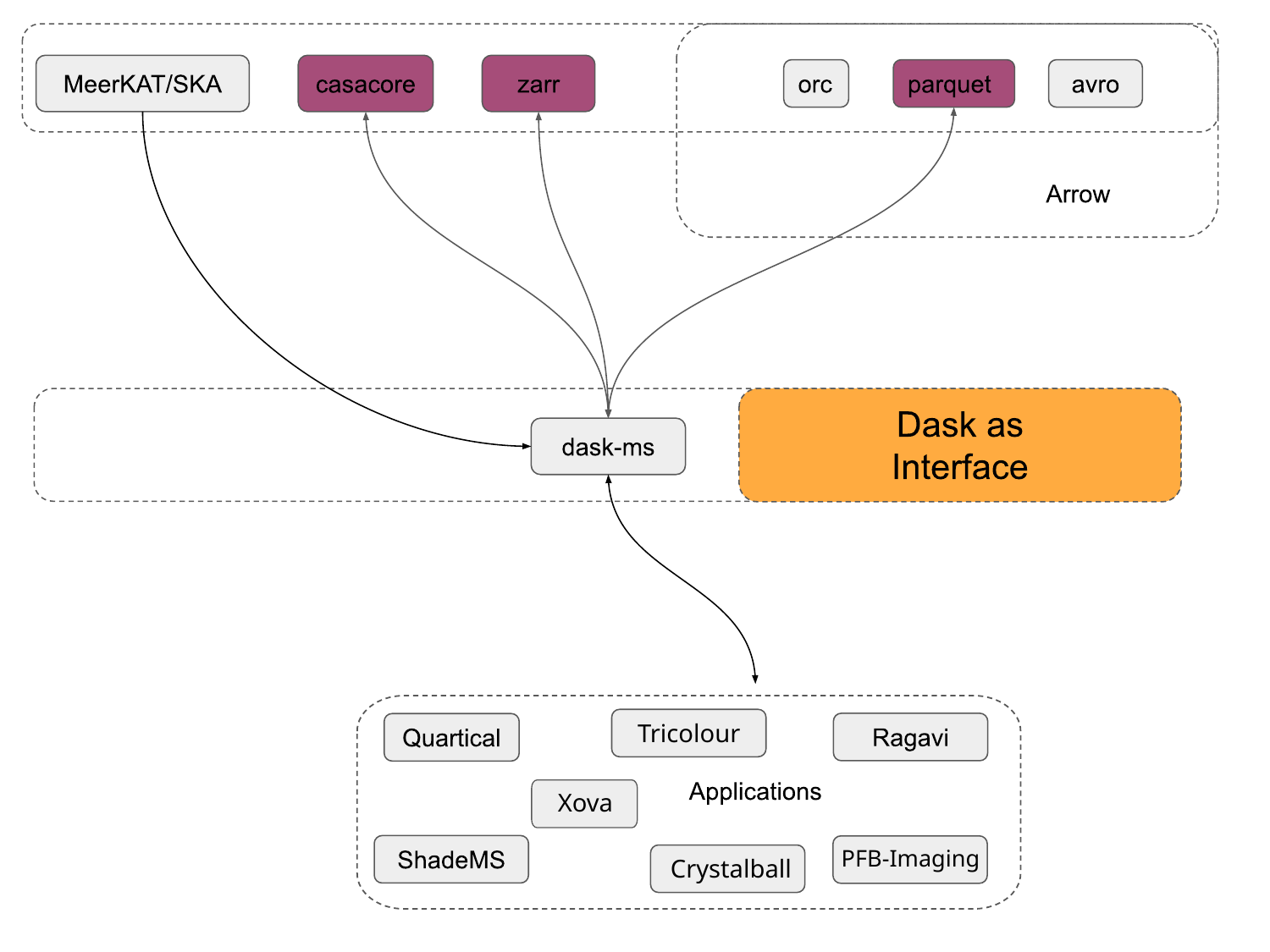}
\end{center}
\caption{dask-ms serves as a Data Access Layer mediating between various data sources and applications.}
\label{fig:daskms-interface}
\end{figure*}

\begin{figure*}
\begin{center}
\includegraphics*[width=0.75\textwidth]{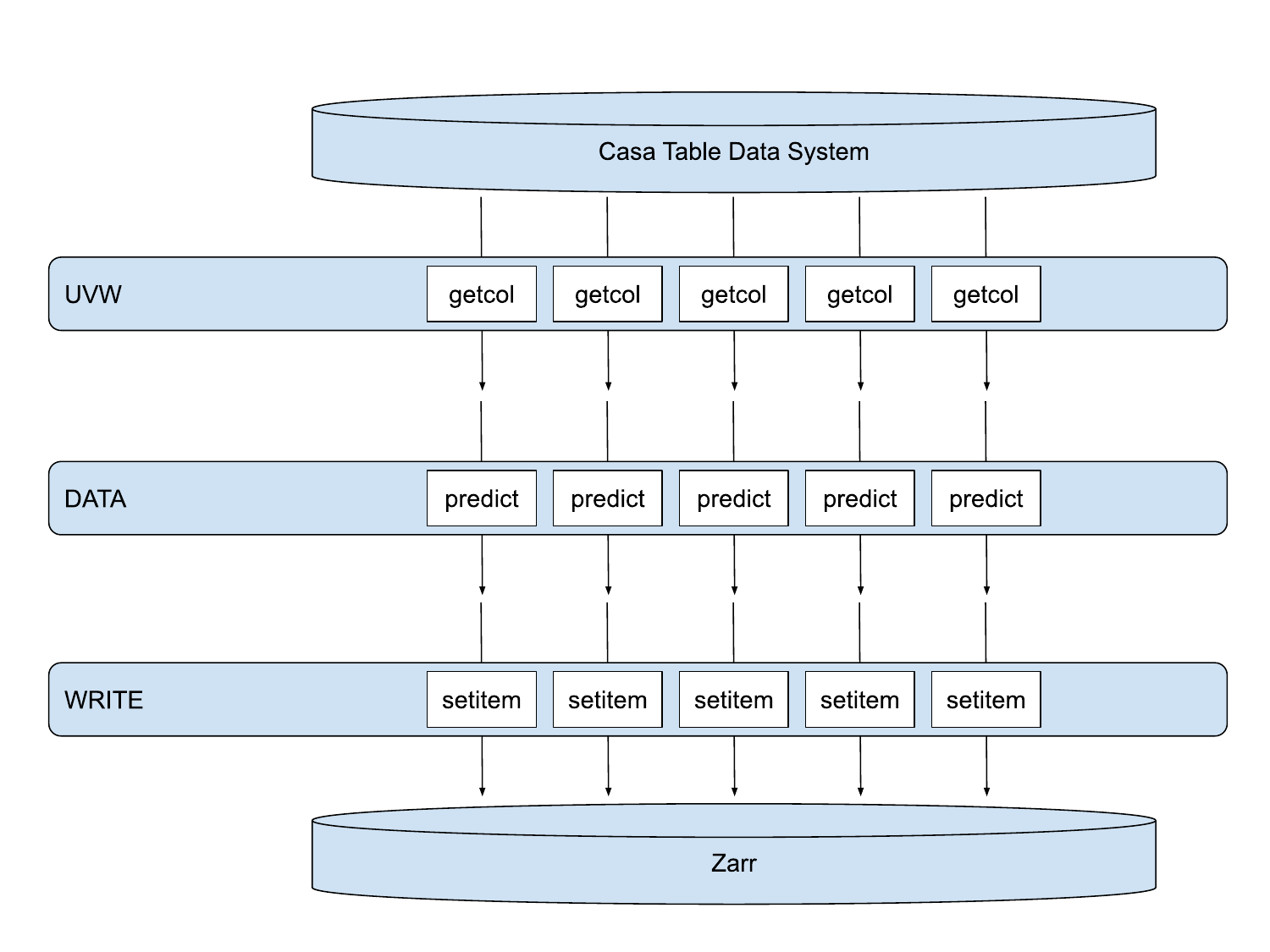}
\end{center}
\caption{Extract, transform, load (ETL) achieved using \dask\ arrays.
UVW, DATA and WRITE are \dask\ arrays that strictly represent {\it loads}, {\it transformations} and {\it extracts} of data, rather than data per se.
Here, the UVW array actually represents five {\it getcol} functions that extract separate
chunks of UVW coordinates.
The DATA array transforms UVW getcol outputs by passing them into individual {\it predict} functions to produce visibilities.
Finally, the WRITE array loads visibilities into a \zarr\ object store via
{\it setitem} operations.
}
\label{fig:dataflow}
\end{figure*}

In the previous section, we described a compelling series of components from the \pydata\ ecosystem. Here we describe the design decisions underpinning a new radio astronomy ecosystem, influenced by our trifecta of desiderata: flexibility, ease-of-development and performance.

{\bf Performance: support multi-core, distributed programming.} As radio astronomy is now firmly in the Big Data regime, any future ecosystem will need to be built on
distributed, multi-core architectures from the ground up, as data volumes are now simply too large for earlier architectures to tractably handle these data quantities.

{\bf Performance: use \numba, \numpy, \scipy\ and C/C++ Extensions.} Many \numpy\ and \scipy\ operations drop the GIL, allowing multi-threaded applications to use multiple cores
in Python.
However, many operations do not have sufficient {\it arithmetic intensity} to fully exercise all cores: elementwise-wise arithmetic operations only perform one FLOP per element of data. Many more FLOPS/byte are needed to exercise modern CPUs \citep{Alted2010}.
Therefore, in critical code sections, we routinely use \numba\ to greatly increase the arithmetic intensity of our algorithms.
In other cases, we can rely on performant code implemented in C/C++ extensions, the $w$-gridder \citep{Arras2021} being a prime example.

{\bf Ease-of-development: enable non-expert software developers to develop new techniques using the \pydata\ ecosystem.}
Both AIPS and CASA were concerted and coordinated software development efforts, well-resourced by the relevant organizations. This is in contrast to the ``bazaar'' development model espoused by the open source community -- we can't think of any substantial bazaar-model projects in radio astronomy. In recent times, the increased capabilities (field-of-view and sensitivity) of our new instruments have driven a lot of  algorithmic innovation, as legacy algorithms are often unable to fully exploit the new science capabilities. The eventual compute requirements of many new algorithms are still poorly understood, which makes it difficult to incorporate them into a formal, top-down software engineering process. This leaves scientific
software developers and astronomers with a large role to play in exploring these algorithms. However, due to the nature of their work, their primary focus is on extracting science results.
To merely prove new techniques with SKA pathfinder telescopes, distributed and multi-core architectures must be taken into account, as data volumes are otherwise too large.
However, scientific developers have little time to devote to low-level implementation concerns. Therefore, it is important to lower the barrier to entry for important new software technologies.
The \pydata\ ecosystem is the obvious choice: most scientific developers and astronomers
know Python, \astropy, \numpy, \pandas\ and \scipy, which provide convenient data structures
and algorithms for developing new techniques, while \dask\ provides distributed programming tools using semantics from \numpy\ and \pandas\ .
{\sc Julia} \citep{Julia-2017} is a modern, powerful language designed for scientific computing with a lot of promise. However, Julia scientific computing libraries are currently less mature and have fewer maintainers than Python.

{\bf Flexibility: decouple algorithms from data sources.} As we are in a regime when many new algorithms are under development, it is important to be able to compose algorithms together. For example, a calibration algorithm may want to ingest model visibilities obtained from component-based or degridding-based prediction algorithms, or a combination of the two, or it may want to obtain model visibilities from data on disk.
For these reasons, we decouple our data sources and sinks from algorithms and applications by implementing a Data Access Layer (DAL) called \daskms\ ($\S$\ref{section:daskms}) and an algorithms library called \codexafricanus\ ($\S$\ref{section:codex}).

{\bf Flexibility: support multiple data sources and sinks.} The decoupling described above also encourages us to separate our algorithms from data sources and sinks.
If we write our algorithms in such a way that they consume and emit \numpy\ arrays,
then the source and destination of these arrays is unimportant.
Data can be obtained from Measurement Sets, other sources on disk, or even other algorithms without
the algorithm having to consider its provenance.
This requirement is naturally provided by lazy \dask\ arrays:
As \dask\ arrays represents a series of operations that generate data,
we can use them to interchangeably represent
(1) reads from multiple disk formats such as the MS, \zarr\ or \arrow\
(2) the output of an algorithm.
In the case of (1), \daskms\ presents \dask\ arrays to the developer without
concerning the developer about the underlying format.
Similarly, in the case of (2), \codexafricanus\ can consume \dask\ arrays produced by
\daskms\, or even other \codexafricanus\ algorithms and produce \dask\ arrays
representing the result of the algorithm.
Finally, when writing data, \daskms\ consumes \dask\ arrays, writing them to them
the underlying format.

{\bf Ease-of-development: use previously defined specifications in new formats and data structures.} As discussed in $\S$\ref{section:background}, the Measurement Set is a {\it specification} for structuring radio interferometry data.
By contrast, the CTDS is an on-disk {\it format} that concretely implements the specification.
The mental model provided by the Measurement Set specification is ubiquitous and ingrained in a generation of developers.
This specification also maps easily to many data structures and formats in the \pydata\ ecosystem.
For example, both \zarr\ and \arrow\ are designed with many of the concepts implemented in the CTDS -- this makes it relatively easy to map the Measurement Set specification onto these new formats.
Similarly, \pandas\ dataframes and \xarray\ datasets collate columnar data into tabular datasets for manipulation by developers.
For this reason, we use the Measurement Set specification in various forms throughout our code.

{\bf Dataflow programming:} Dataflow programming is a suitable method for achieving most of these aims.
By representing programs as a graph, task schedulers can distribute work on multiple nodes and cores.
\dask\ arrays provide a familiar and convenient mechanism for defining dataflow graphs as they semantically behave like \numpy\ arrays. To the scientific developer it appears that they are composing a series of \numpy\ -like objects together, while in reality this iteratively builds a dataflow graph.
This graph is then submitted to a scheduler for execution on multiple nodes and cores.
The operations in such dataflow graphs are loosely-coupled: an algorithm data source can be swapped simply by substituting a different \dask\ array as input during graph construction.
For example, Figure \ref{fig:dataflow} shows how \dask\ arrays can be used to represent {\it Extraction, Transformation and Load} (ETL) \cite{Kimball2004} data processes.
This style of programming makes code built in this paradigm easy to read and maintain.

\section{\daskms\ }
\label{section:daskms}

\daskms\ is a {\it Data Access Layer} written as a Python package that exposes Measurement Set-like data as \xarray\ datasets containing \dask\ arrays.
We say they are Measurement Set-\emph{like} because the interface \xarray\ presents to
the scientific software developer is very similar to the tables and columns of an MS.
It provides a familiar basis for writing distributed \pydata\ radio astronomy applications, as shown in
Figure~\ref{fig:daskms-interface}.

\dask\ arrays produced by a read function (\mintinline{python}{xds_from_ms}) represent a chunked read of on-disk data for that column.
This data can be stored in a \casa\ Measurement Set, but could also be stored in \zarr\ or \arrow\ datasets.
Conversely, \dask\ arrays produced by a write function (\mintinline{python}{xds_to_table}) encode a series of write operations.
These arrays are actually lazy graphs: no operation takes place until the graph is explicitly evaluated.

\xarray\ logically groups \dask\ arrays together and provides the developer with a convenient structure for manipulating data, as shown in Listing \ref{listing:xarray}.
Dimensions can be labelled with familiar names, such as row, channel and correlation in the case of Measurement Set data.
Coordinates can be assigned along each dimension
allowing arrays to aligned against each other and for ranges of data to be manipulated by quantities such as time and frequency.
\daskms\ does not explicitly assign coordinates, as row dimension coordinates could grow to be very large, but application developers are free to assign these coordinates to the \xarray\ Datasets themselves.

Measurement Set DATA columns can be variably shaped, which is unsupported by \dask\ arrays.
To work around this, a Measurement Set is partitioned into multiple \xarray\ datasets (primarily by DATA\_DESC\_ID and FIELD\_ID values), thereby ensuring each dataset has a homogeneous channel and correlation shape

\daskms\ has strong support for \casa\ tables. It supports read, create, update and append operations to \casa\ tables. Read operations simply produce a list of \xarray\ datasets. Create operations accept a list of \xarray\ datasets and create an entirely new dataset, inferring the table and column descriptors, as well as data managers for each column. If columns share the same shape across each dataset, they will be assigned Tiled Storage Managers.

Update operations allow the modification of existing columns, and the addition of new columns to the dataset. A special ROWID coordinate column must be present on the written \xarray\ dataset to indicate support updates back to the Measurement Set (read operations produce it).
Finally, append operations allow data to be added to a table, although this case is usually only associated with the creation of a new table.

\daskms\ also supports the distributed, cloud-native \zarr\ and \arrow\ formats.
Measurement Set data can be written in both formats, with some caveats.
As \zarr\ represents multi-dimensional arrays, variably shaped-data must be written as a series of datasets.
This is achieved by using TAQL \citep{VanDiepen2006} to partition the Measurement Set by it's key columns.
The resulting \zarr\ stores can be read, created, updated and appended to.

By contrast, \arrow\ can represent variably-shaped data, but only supports chunking along the primary row dimension, although this limitation could be mitigated by stacking datasets along other dimensions (channel being the obvious one). It only supports reading and writing: updating existing data is not supported. It is possible to append to an existing dataset by adding parquet files into the existing directory structure.

While the above formats differ in the underlying storage structure, the Measurement Set v2.0 specification is always maintained as far as the application code interface is concerned.

{\bf \katdal\ export to \zarr\ Measurement Sets:} \katdal\ \citep{Schwardt2023} is the Data Access Layer to the MeerKAT Data archive.
It provides its own Dataset interface over archived observations, allowing selection along
multiple dimensions including time, frequency, spectral window and antenna.
It also provides export of archived observations into the CASA Measurement Set v2.0 format.
Due to the limitations of the CTDS as a distributed computing format,
we have implemented export from \katdal\ into an Measurement Set v2.0 format backed by \zarr\ on disk.
As \daskms\ can seamlessly operate with CTDS or \zarr\ formats, this provides
applications constructed on top of \daskms\ with the ability to run on a fast, distributed
storage system.

\section{\codexafricanus\ }
\label{section:codex}

\codexafricanus\ \footnote{\url{https://github.com/ska-sa/codex-africanus}} is a Python \textit{Application Programming Interface} (API) which exposes radio astronomy algorithms as Python functions.
The purpose of \codexafricanus\ is to facilitate the re-use of algorithms within radio astronomy applications.
For example, a model prediction algorithm is useful in both calibration and imaging, but to re-implement it in separate contexts is wasteful.
Furthermore, different prediction algorithms exist: component-based (a.k.a. DFT) predicts are
computationally expensive, but their accuracy is desirable for bright sources, while the fast but less accurate FFT-and-degrid approach is more suitable for fainter extended emission. Both strategies can be usefully combined in the same application.
\codexafricanus\ aims to meet these requirements.

Two layers are involved: in the first, a function ingesting and outputting \numpy\ arrays is defined to implement the algorithm on a \textit{chunk} of input.
For performance, these functions are frequently implemented in \numba, to obtain efficient machine-code with performance
characteristics similar to C++ and Fortran.
The second layer is a function that defines a transformation which ingests and outputs \dask\ arrays.
This concept is demonstrated in Figure \ref{fig:dataflow}, where the DATA array wraps a number of predict function calls that transform the individual chunks of the UVW array.

Amongst others, the following notable algorithms are exposed:

\begin{enumerate}
    \item Component-based model predict with an assortment of Jones matrices, following the radio interferometer measurement equation \citep[RIME;][]{Smirnov2011} formalism.
    \item The $w$-gridder \citep{Arras2021}, which performs gridding/degridding of visibility data;
    \item A Perley-style polyhedron faceting gridder and degridder \citep{Cornwell1992}.
    \item Baseline-dependent time and channel averaging (BDA) \citep{Atemkeng2016}.
\end{enumerate}

The flexibility and speed of \numba\ is exhibited in the \codexafricanus\ RIME implementation.  A string defining the RIME for a series of terms is provided
to a \numba\ interface function.
For example the string:
\begin{equation}
\textrm{[Ep, Lp, Kpq, Bpq, Lq, Eq]: [I, Q, U, V] -> [XX, XY, YX, YY]}
\end{equation}
defines a RIME with per-antenna beam and feed terms as well as
per-baseline phase and brightness terms, that transforms the supplied
Stokes values into the relevant correlations
$p$ and $q$ refer to left and right per-antenna terms which must be
present on the left and right side of the equation, respectively. $pq$ indicates
a per-baseline term which must be present in the middle of the equation.
Each element in this string is identified (terms, stokes and correlations),
used to {\it dispatch} functionality to different Jones term implementations
and construct a single JIT-compiled function that multiplies out the Jones chain
in an inner loop to form source coherencies.
This is a powerful example of using {\it multiple dispatch} where the function
for computing one class of RIME is defined by a string argument.

\section{Dask Scheduling}
\label{sec:dask_scheduling}

One of \dask's primary strengths is the expressiveness afforded to the scientific software developer by \dask\ Dataframe and \dask\ Array collections.
As they have similar semantics to \numpy\ arrays, they provide a familiar {\it affordance} \citep{Norman2002}, providing developers with the ability to rapidly develop distributed programs.
The tradeoff with this abstraction is that it transfers the responsibility of {\it scheduling} the graph from the developer to \dask's schedulers, which may -- or may not! -- optimally assign tasks to nodes.

In particular, {\it spatial locality} is an important factor in any distributed application (or multi-core application).
Related pieces of data should be co-located on the same cluster node in order to avoid the overhead of
network transfers slower than that of a main memory bus.
This is especially important in the context of an \xarray\ dataset of \dask\ arrays whic are chunks of related arrays
should ideally be co-located.
For example, in the context of the Measurement Set, the DATA, WEIGHT\_SPECTRUM and FLAG columns have the same shapes,
are all very large and their related chunks should be co-located to avoid data transfers.

This can be a difficult problem: optimally scheduling task graphs
with arbitrary constraints is NP-hard \citep{Ullman1975} and
heuristic solutions are needed in practice \citep{Kwok1999}.
\daskdistributed\ applies a number of best effort heuristics \citep{DaskDistributedScheduling1999},
noting that correct initial placement of tasks is an important factor.

With the intention of ensuring correct placement,
{\it Task Annotations} were added to \dask\ \citep{DaskLayerAnnotations2020}
allowing arbitrary metadata to be associated with \dask\ tasks.
This metadata is passed through to \daskdistributed\ Scheduler plugins,
providing hints to the scheduler that can be used to optimise task placement.

Consider, for example, the WEIGHT\_SPECTRUM and DATA columns,
represented with \dask\ Arrays having the same shape and chunks.
We would like to ensure that related row chunks are co-located on the same
\dask\ worker.
One method of achieving co-location is to take advantage of \dask's
naming convention for Array collection graph keys, which are tuples of the form
$(\textit{name}, {i_1}, ..., {i_n})$ where $i_{n}$ represent
the integral chunk indices of each dimension and $\textit{name}$ is a unique
name identifying a collection.
Then, the \lstinline{WEIGHT_SPECTRUM} and \lstinline{DATA} \dask\ arrays might have unique
names \lstinline{WEIGHT_SPECTRUM-a2b4} and \lstinline{DATA-3fc6}.
We also know they share the same \lstinline{(row, chan, corr)}
dimensions and we would therefore like to schedule two related chunks,
uniquely identified by keys \lstinline{(WEIGHT_SPECTRUM-a2bf, 3, 1, 0)} and
\lstinline{(DATA-3fc6, 3, 1, 0)}, on the same worker.
It is relatively simple to annotate each task with
a tuple \lstinline{(row,chan,corr), (10, 16, 4))}.
containing both the dimension name and the number of chunks in the dimension.
Then, it is simple to (1) infer that each task is associated with row chunk 3 (out of 10) and (2) stripe the task over a range of workers

\begin{equation}
\textrm{worker\_id} = \textrm{floor}(\textrm{nworkers} \times \textrm{rowchunk\_id} / \textrm{nrowchunks})
\end{equation}

If the developer is developing using \xarray\ Datasets,
this dimension name metadata is automatically associated with \dask\ Arrays.
Pseudo-code for the above strategy is shown in Listing~\ref{listing:scheduler_plugin}.

\begin{listing*}[!ht]
\begin{minted}{python}
# Create a chunked dask array and then clone it with a
# name_chunks annotation for each key
DATA = dask.array.ones((1000, 64, 4), chunks=(100, 16, 4))
with dask.annotate(name_chunks=(("row", "chan", "corr"), A.numblocks)):
    # Cheaply recreate A with annotations
    DATA = dask.graph_manipulation.clone(A)

class RowStripePlugin(SchedulerPlugin):
    """ Stripes tasks containing row chunks across workers.
    annotations have the form {annotation_name: {key: value}}"""
    def update_graph(self, scheduler, keys, annotations=None, **kwargs):
        worker_names = list(scheduler.workers.keys())
        nworkers = len(worker_names)
        # Returns the dimension id of the row dimension
        # given an iterable of the form (dim_id, dim_name)
        find_row_id = lambda i, d: d[0] if d[1] == "row" else i

        for key, (names, chunks) in annotations["name_chunks"].items():
            if row_id := reduce(find_row_id, enumerate(names), None)
                # Collection keys have the form (name, i, j, k)
                assert isinstance(key[0], str) # Collection name in first position
                row_chunk = key[row_id + 1]    # Get the row chunk from collection key
                row_chunks = chunks[row_id]    # Number of row chunks

                if ts := scheduler.tasks.get(key):
                    worker_id = math.floor(nworkers * (row_chunk / row_chunks))
                    ts.worker_restrictions = {worker_names[worker_id]}
                    ts.loose_restrictions = True
\end{minted}
\caption{A simple scheduler plugin that stripes row chunks over
available dask workers.
Arrays are annotated with dimension names
and the total number of chunks in each dimension.
The plugin uses this metadata to determine the row chunk id
for a particular task and stripe the task over available workers.}
\label{listing:scheduler_plugin}
\end{listing*}

\section{Results: An Ecosystem of radio astronomy Applications}
\label{section:result_ecosystem}

The flagship applications in the Africanus ecosystem are
\QuartiCal\ \citep{africanus2} and \pfbimager\ \citep{africanus3}, as they are fairly
large and rich-featured packages exemplifying the design criteria described
in $\S$\ref{section:ecosystem}. They will be covered separately in the following papers of this series.
In this section, we describe a number of smaller packages developed within the ecosystem. Taken as a whole, this suite of packages is sufficiently feature-complete to support (at least) conventional data reduction workflows, from raw visibilities to final images. Note that some of these packages (notably, \tricolour, \shadems, \ragavi, {\sc crystalball} and \QuartiCal) have already found use in end-to-end pipelines such as {\sc CARACal} \citep{caracal}, while others such as {\sc Tabascal} and {\sc GridFlag} have been developed under completely independent authorship.

\subsection{\tricolour}

\tricolour\ \citep{Hugo2022} is a \dask-based radio frequency interference (RFI) flagging application implemented in \dask\ and \numba.
It implements the SumThreshold \citep{offringa2012} algorithm employed in the AOFlagger \citep{AOFlagger2010} package, but is capable of efficiently handling the quantities of data produced by MeerKAT in its 32k channel mode.

The SumThreshold algorithm detects RFI by applying multiple filter-passes to per-baseline time and channel windows.  In the case of Measurement Set-like data, we perform a transpose from (row, channel, corr) ordering to (baseline, corr, time, channel) {\it per scan}, allowing time-channel windows to be flagged in parallel.
This transpose can be performed in memory, or via on-disk \zarr\ arrays.

\tricolour\ is multithreaded but not distributed. It scales well to $\sim$20 CPU cores and, given this number of cores, flags data at over 400 GiB/h \citep{Hugo2022}.

\subsection{\shadems}

It is useful to visualize the raw (or calibrated) visibilities produced by an interferometer, as this can often reveal subtle systematics, and yield insights into the quality of the calibration. The go-to tool for this has been the CASA \plotms\ task, which uses the {\sc matplotlib} \citep{matplotlib2007} library under the hood, mainly in scatter plot mode. This has a hard limit of just over 4 billion datapoints, and becomes very inefficient (and completely saturates the plot canvas) long before this limit is reached. Since the data produced by modern interferometers easily exceeds this limit, \plotms\ invocations are typically restricted to carefully chosen subsets (or averaged-down versions) of the data.

\datashader\ \citep{Datashader2022} is a recent plotting package fully integrated with \dask.
Portions of the data (\dask\ array chunks) are rendered onto separate canvases and aggregated to form a final plot. Crucially, it uses \emph{shading} rather than plotting: pixels of the canvas are assigned colour/alpha values that represent the density of the data points falling therein (rather than rendering each data point as an invididual marker). This avoids the saturation problem, and means that any size of dataset can be rendered meanigfully. Since multiple chunks can be processed on multiple cores and nodes, this also greatly accelerates canvas rendering. \shadems\ \citep{shadems2022} is a tool for rapidly plotting Measurement Set data that uses \datashader\ to render \dask\ arrays produced by \daskms.
It achieves considerable acceleration -- as a qualitative indication, a 32-core machine will render ~4 billion points in under 3 minutes, compared to over an hour for \plotms\ to plot an $\times8$ averaged-down version of the same data.

\subsection{\ragavi\ }
The radio astronomy gains and visibilities plotter \citep[\ragavi,][]{Ragavi2022} generates interactive plots of calibration solutions and visibilities for radio astronomical data. It uses \daskms\ as a data access interface between the data processing backends, \dask\ and \numpy, and plotting software -- namely \textsc{Bokeh, Matplotlib}, and \textsc{Datashader}.\footnote{\url{https://bokeh.org/}} The backend and plotter choices depend on the input data size during runtime. Typical calibration solutions are small in size (within a few megabytes), containing less than 5,000 data points, and can easily be contained and processed in RAM without drastic memory consumption, thus permitting the use of \numpy\ for data crunching. Furthermore, it is possible to plot each data point in this scenario, enabling the level of interactivity afforded by \textsc{Bokeh} plots (which ultimately generates interactive HTML documents). When data sizes increase to visibility scale, \ragavi\ falls back on a more \shadems-like strategy that uses \datashader\ to render static plots.

\subsection{Others}

{\sc xova} \citep{atemkeng2021} is an averaging application that performs both time and channel averaging and baseline dependent averaging along either or both axes.
In the latter case, each averaging bin is assigned a unique TIME and INTERVAL that limits smearing to an acceptable decorrelation tolerance.
A unique CHAN\_FREQ and CHAN\_WIDTH is not realistically achievable in the context of the Measurement Set v2.0 specification as, in the general case, this would
require a separate spectral window per row. As a workaround, the number of channels in the original spectral window is factorised by increasing powers of two, and new spectral windows are created from the resulting subdivisions. The loss in compression is negligible, as shown in \citep{Hugo2024}. Both algorithms are exposed in \codexafricanus.

{\sc crystalball} predicts the visibilities corresponding to a component source model using a DFT-based predict implemented in \numba. The current model format is restricted to a \wsclean\ source list,  but other model formats could be added in the future.
The predict itself is parallelised over {\bf row}, {\bf channel} dimensions and {\bf source} dimensions.
The DFT is an extremely computationally demanding, but accurate, method of predicting model visibilities from
a source model.
As such, it is well-suited to demonstrating the strong and weak-scaling properties of a distributed framework,
as demonstrated in $\S$\ref{section:results_dist_predict}.

The {\sc Gridflag} algorithm  \citep{Sekhar2018}\footnote{\url{https://github.com/idia-astro/gridflag/}} aggregates residual visibilities
into a UV-cell bin, based on their UV-coordinates.
Gridflag assumes that residuals should differ by system temperature: residuals outside this range are discarded.
Internally, Gridflag parallelises this processing by reading a chunked residual visibility column from a CASA Measurement Set using \daskms, while performing UV-binning and flagging using \numba\ kernels.

{\sc Tabascal} \citep{Finlay2023} \footnote{\url{https://github.com/chrisfinlay/tabascal}} jointly models calibration and RFI parameters
for satellites with known trajectories.
Internally, it uses \daskms\ and \jax\ \citep{jax2018} to distribute compute over multiple CPUs and GPUs.

Last, but not least, the {\sc DDFacet} package \citep{ddfacet2018} has recently been extended to use \daskms\ as an alternative I/O layer, making it ultimately compatible with our new ecosystem.

\section{Results: distributed DFT predict on the cloud}
\label{section:results_dist_predict}

The {\sc crystalball} package mentioned previously implements a DFT-based model predict from a source component list into visibilities. This operation is commonly done in the major cycle of imaging, as well as when simulating data. However, due to its computational expense (scaling linearly with the number of model components and shape of the visibilities, which can be large), it is often implemented using pixellated model images, via an FFT-and-degrid cycle (scaling as $O(N_{\mathrm{pix}}\log N_{\mathrm{pix}}$)), at a reduced frequency resolution.

There are important use cases when the expense of a full DFT predict is justified. These include high-dynamic range imaging, as well as continuum subtraction in spectral line imaging, where it is important to predict the brighter model components  at the highest precision and frequency resolution available. {\sc crystalball} was developed to meet these needs (and is routinely used in the {\sc CARACal} pipeline for continuum subtraction).

In the simplest case, the DFT predict is just a sum over model components $s$:
\begin{equation}
V_{pq}(t,\nu) = \sum_{s} I_s(\nu) \times \mathrm{e}^{-2\pi i(u_{pq}(t)l_s + v_{pq}(t)m_s + w_{pq}(t)(n_s-1))}
\end{equation}
where $pq$ is the baseline index, $V$ is the visibility, $u$, $v$ and $w$ are the baseline coordinates in units of wavelength, $I_s(\nu)$ is the component flux as a function of frequency, and $l_s, m_s$ are the component coordinates. Since baselines rotate in time, the $uvw$ coordinates are a function of both time and frequency.

In addition to the baseline dimension, the DFT predict must be independently evaluated at distinct times and frequencies. The independence of these dimensions means the DFT is highly parallel, making it amenable to GPUs \citep{Perkins2015} and distributed computation.

\codexafricanus\ contains a \numba\ implementation of the DFT predict. While some I/O is required to read in UVW coordinates and write out the visibilities, the predict is dominated by computation, and we would therefore expect to achieve both strong and weak scaling, as described both by Amdahl and Gustafson's laws.
The following experiments investigate whether this is achievable with \dask\ and \numba, while running the predict in a cloud environment, using the AWS EKS implementation.



{\bf Strong scaling:} for our test case, we use a MeerKAT L-band observation of ESO137-006 \citep{Ramatsoku2020}. This contains 3149 timesteps (at 8s dump time), 1891 baselines (61 antennas) including auto-correlations, for a total of 5,954,759 rows. Each row contains 4096 channels and four correlations yielding a total of 345GB of visibility data.

This dataset was converted into \zarr\ format and split into chunks of 50,000 rows. For our test run,
we predict 100 point source components (corresponding to 34TB of predicted source coherencies, prior to summing over $s$). The \dask\ computation consisted of 10388 tasks. We used Amazon {\tt m5.4xlarge} EC2 instances with 16 vCPUs and 64GBs of RAM each.

\begin{table*}
\centering
\begin{tabular}{|c|c|c|c|c|}
\hline
{\tt m5.4xlarge} count & Wall time (seconds) & Aggregate Clock Time (minutes) & Average Tasks/thread\\
\hline
10 & 934 & 2372 & 65 \\
20 & 490 & 2285 & 32 \\
30 & 349 & 2271 & 22 \\
40 & 271 & 2083 & 16 \\
50 & 215 & 2157 & 13 \\
80 & 145 & 2136 & 8.1 \\
100 & 134 & 2198 & 6.5 \\
\hline
\end{tabular}
\caption{Strong scaling results showing wall time and aggregate clock time for increasing numbers of m5.4xlarge EC2 instances. The wall time represents the amount of time taken for that iteration of the experiment to run, while the aggregate clock time is the sum of time spent computing across all instances. The number of tasks remained constant at 10388 for all runs}
\label{tab:strong_scaling}
\end{table*}

\begin{table*}
\centering
\begin{tabular}{|c|c|c|c|c|}
\hline
m5.4xlarge count & Wall time (seconds) & Aggregate Clock Time (hours) & Average Tasks/threads & Total Tasks\\
\hline
10 & 936 & 38.6 & 65 & 10388 \\
20 & 1126 & 88 & 59 & 18935 \\ 
30 & 1328 & 159 & 62  & 29900 \\
40 & 1549 & 237 & 57 & 43283 \\
50 & 1766 & 349 & 85 & 68027 \\ 
80 & 2512 & 592 & 104 &  133025 \\ 
100 & 2975 & 540 & 123 & 196200 \\
\hline
\end{tabular}
\caption{Weak scaling results showing wall time and aggregate clock times for increasing numbers of m5.4xlarge EC2 instances and tasks}
\label{tab:weak_scaling}
\end{table*}

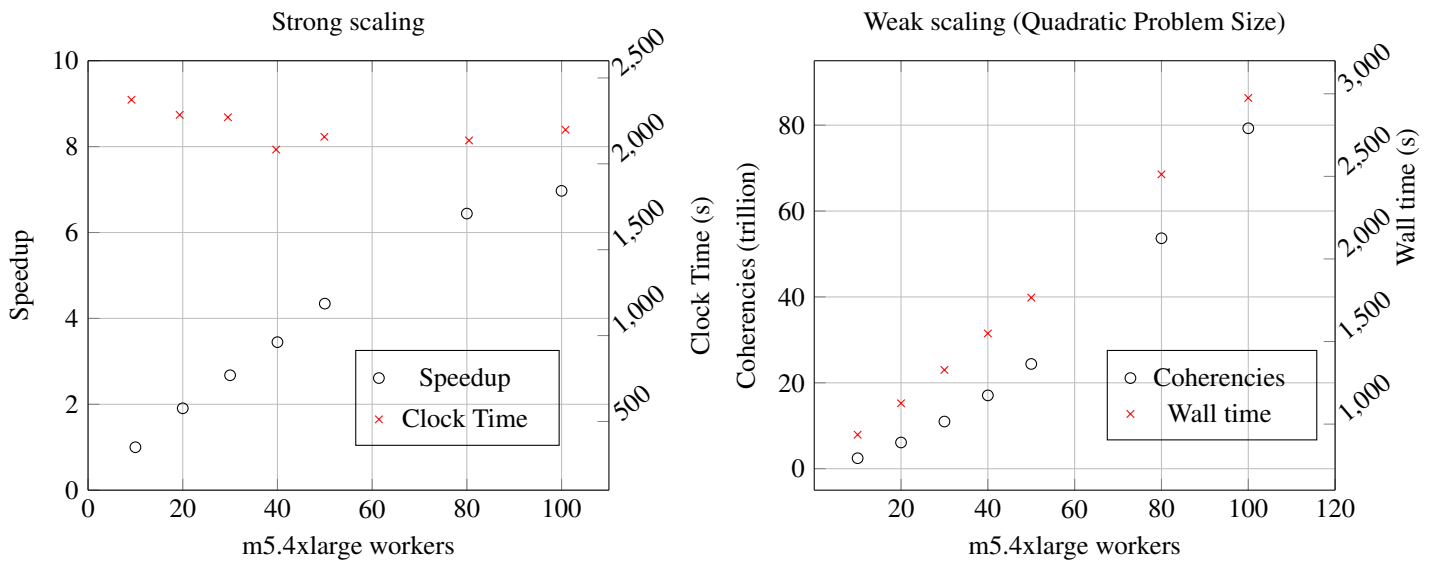
\begin{figure*}
\begin{tikzpicture}
\hypersetup{hidelinks}
\begin{axis}[
    title={Strong scaling},
    name=ax1,
    xlabel={m5.4xlarge workers},
    ylabel={Speedup},
    xmin=0,
    ymin=0, ymax=10,
    axis y line*=left,
    grid=major,
]
\addplot[only marks, mark=o,black] coordinates {
    (10, 934/934)
    (20, 934/490)
    (30, 934/349)
    (40, 934/271)
    (50, 934/215)
    (80, 934/145)
    (100, 934/134)
};
\label{fig:strong_scale_speedup}
\end{axis}
\begin{axis}[
    axis y line*=right,
    axis x line=none,
    ylabel={Clock Time (s)},
    yticklabel style={
        rotate=40,          
        anchor=west,        
        /pgf/number format/.cd,
        scaled y ticks = false
    },
    ymin=100, ymax=2600,
]
\addplot[only marks, mark=x,red] coordinates {
    (10, 2373)
    (20, 2285)
    (30, 2271)
    (40, 2083)
    (50, 2157)
    (80, 2136)
    (100, 2198)
};
\label{fig:strong_scale_time}
\end{axis}
\matrix[
    draw,
    matrix of nodes,
    anchor=south east,
    below=1mm,
] at ([xshift=-20mm, yshift=30mm]ax1.outer south east) {
    \ref{fig:strong_scale_speedup} & Speedup &[5pt] \\
    \ref{fig:strong_scale_time} & Clock Time \\
};
\end{tikzpicture}
\label{fig:strong_scaling}
\begin{tikzpicture}
\hypersetup{hidelinks}
\begin{axis}[
    title={Weak scaling (Quadratic Problem Size)},
    name=ax1,
    xlabel={m5.4xlarge workers},
    ylabel={Coherencies (trillion)},
    xtick={20, 40, 60 ,80, 100, 120},
    xmin=0, xmax=120,
    ymin=-5,
    ymax=95,
    axis y line*=left,
    grid=major,
]
\addplot[only marks, mark=o,black] coordinates {

    (10,  2.44)
    (20, 6.1)
    (30, 10.98)
    (40, 17.07)
    (50, 24.39)
    (80, 53.66)
    (100, 79.27)


};
\label{fig:weak_scaling_coh}
\end{axis}
\begin{axis}[
    axis y line*=right,
    axis x line=none,
    ylabel={Wall time (s)},
    ylabel near ticks,
    xmin=0, xmax=120,
    y label style={anchor=west},
    yticklabel style={
        rotate=40,          
        anchor=west,        
        /pgf/number format/.cd,
        scaled y ticks = false
    },
    ymin=600, ymax=3200,
]
\addplot[only marks, mark=x,red] coordinates {
    (10, 936)
    (20, 1126)
    (30, 1328)
    (40, 1549)
    (50, 1766)
    (80, 2512)   
    (100, 2975)
};
\label{fig:weak_scaling_time}
\end{axis}


\matrix[
    draw,
    matrix of nodes,
    anchor=south east,
    below=1mm,
] at ([xshift=-20mm, yshift=30mm]ax1.outer south east) {
    \ref{fig:weak_scaling_coh} & Coherencies &[5pt] \\
    \ref{fig:weak_scaling_time} & Wall time \\
};
\end{tikzpicture}
\caption{Strong and weak scaling plots for Table \ref{tab:strong_scaling} and \ref{tab:weak_scaling}. In the strong scaling case, the aggregate clock time is also plotted to show how it stays relatively constant, which should be the case if the problem size is held constant. In the weak scaling case, the Wall time is plotted to show how increasing the amount of work per worker commensurately increases the Wall time.}
\label{fig:scaling}
\end{figure*}

The results of this experiment, extracted from \dask\ performance reports\footnote{Available at \url{https://github.com/sjperkins/predict}}, are presented in Table~\ref{tab:strong_scaling} and Figure~\ref{fig:scaling}. Aggregate clock time, summed over all instances, remains relatively constant (2080--2400 seconds across all instance counts). Increasing the instance count from 10 to 100 {\tt m5.4xlarge} instances produces a linear speedup, showing that strong scaling holds until about 80 workers, when the benefits of adding more workers start to fall off. The reason for this is that, at 80 workers, the {\it average} number of tasks per thread drops to a critical point (8.1, at this stage), negating the speedup obtainable by assigning further workers to the problem. Simply speaking, the size of the problem (78TB of source coherencies) is too small to efficiently solve with further parallelism.

{\bf Weak Scaling:} Rather than keeping the problem size fixed, Gustafson's law examines whether workloads scale proportionately to the number of processors (instances) assigned to the problem. For this experiment, we keep the ``row'' dimension of the dataset constant while scaling the ``channel'' and ``source'' dimensions by $(3072 + (1024 \times W))$ and $(100 \times W)$ respectively, where $W$ is the number of workers.\footnote{CHAN\_FREQ and source models are generated as on-the-fly \dask\ arrays}

The results of this experiment can be seen in Table \ref{tab:weak_scaling} and Figure \ref{fig:scaling}. As the problem size (measured in number of coherencies computed) quadratically increases per worker, so does the wall time in seconds, demonstrating weak scaling for the predict algorithm.
Indeed, while our problem sizes grow far larger than those in the strong scaling experiment,
no breakdown in scaling occurs because each worker is assigned more than sufficient work.

Granted, a DFT-model predict is one of the simpler algorithms in the \codexafricanus\ toolkit, with the most benign scaling properties. Papers II and III of the series will demonstrate scaling of far more elaborate algorithmic structures. The point of this experiment was to demonstrate \emph{extreme} scaling of a computationally-bound algorithm on a commodity compute platform, thus showing that our underlying \dask\ and \numba\ technology choices do not restrict (and, in fact, enable) massive scaling.

\section{Discussion}

Through the use of (1) open source software packages in the \pydata\ ecosystem
and (2) the application of the design criteria described in $\S$\ref{section:ecosystem},
an ecosystem of radio astronomy software, capable (in principle) of processing the large quantities of data produced  by modern interferometers, has been developed.
The flagship packages of this ecosystem, \QuartiCal\ and \pfbimager,
are capable of running both on supercomputers and cloud computing platforms,
as will be discussed in subsequent papers in this series.

The use of open source packages has greatly accelerated development
of our radio astronomy applications.
Indeed, the use of \numpy\ and \scipy\ is nowadays so ubiquitous
that it deserves little discussion, but only warrants further praise
for the heroic efforts of open source contributors.
Furthermore, the use of \numba\ and \dask\ has produced speedups and
distributed applications that, in a more traditional approach,
would have required expert C++ and distributed computing knowledge. This
merits some further discussion.

\subsection{\numba}
\numba\ is extremely convenient for both accelerating and distributing code.
This convenience and speed is primarily achieved through
its support for (1) JIT-compilation and (2) {\it multiple dispatch}.
Achieving similar speed-up the traditional way would typically
require writing a C or C++ extension that interfaces with Python via a library
such as PyBind11 \citep{pybind11}, with additional effort then required to
build binary wheels for distribution across multiple operating systems and
hardware architectures. This is a laborious undertaking.

By contrast, \numba\ calls the LLVM compiler
via the {\sc llvmlite}\footnote{\url{https://github.com/numba/llvmlite}} wheel.
This means that an application effectively needs
to install a compiler to access \numba's functionality, but we consider this
an acceptable trade-off given development speed, flexibility,
performance, and avoiding creation of binary wheels
for our applications (with the attendant packaging concerns).

\subsection{\dask}

{\bf \dask\ is simple to install and run.}
\dask\ is a lightweight Python library with no binary wheel dependencies.
This makes it easy to install in diverse environments such as local laptop environments, supercomputers and
cloud compute environments.
The distributed version provides an informative dashboard
providing amongst other views (1) a task stream showing a short
history of tasks running on each worker (2) a tab showing
CPU and memory usage for each worker and
(3) a statistical profiler showing the amount of time taken
for each distributed task.

{\bf \dask\ Collections provide compelling Interfaces.}
\dask\ offers powerful abstractions, such as the \dask\ Array collection.
In terms of Human Computer Interaction (HCI), these {\it affordances} \citep{Norman2002}
are appealing to the Scientific Software Developer as they build on existing paradigms
such as the \numpy\ CPU array, \cupy\ \footnote{\url{https://cupy.dev/}} GPU array,
the \pandas\ CPU Dataframe and the \cudf\ GPU Dataframe \footnote{\url{https://docs.rapids.ai/api/cudf/stable/}}.
These collections can be provided to software developers by a data access layer
such as \daskms\ for consumption by downstream applications, enabling rapid development of applications.

In our experience, such applications generally perform well on single nodes when used with the \dask\ threaded and process schedulers.
More care is needed in the case of the \daskdistributed\ scheduler which has the added constraint
of managing data transfer between cluster nodes over a network.
We have not evaluated \dask\ with CUDA-based libraries such as \cupy\ and \cudf, but
support for these libraries is advertised at \url{https://docs.dask.org/en/stable/gpu.html}.

{\bf Graph Complexity.} In our experience, \dask\ does well with relatively simple graphs:
parallel, independent streams of execution are the easiest for the schedulers to process
as this correspond to a depth-first traversal of a graph.
The introduction of more complex task dependencies is challenging for schedulers, as large
task outputs may need to be retained until (1) they are consumed by all dependent tasks and
(2) all other dependent task inputs are ready.
Broadly speaking, complex tasks can introduce breadth into the graph, which
tends to increase the amount of memory required for processing the graph.

{\bf Distributed Collection Scheduling.} An important challenge for the \daskdistributed\ scheduler is
initial task placement \citep{DaskDistributedScheduling1999}.
The scheduler assigns tasks to workers without taking the full graph topology,
or the spatial locality encoded into \dask\ collections, into account.
Ignoring these properties results in
(1) memory pressure when results are stored in sub-optimal locations
and (2) network transfers which tend to be much slower than main memory.
We have attempted to ameliorate this by annotating collections with hints
that \emph{scheduler plugins} can use to enforce spatial locality, as described
in $\S$\ref{sec:dask_scheduling}.
This can work well for applications with relatively simple graphs.
QuartiCal \citep{africanus2} is an example of a distributed application
written entirely in terms of collections, which uses a scheduler plugin
to ensure spatial locality.

However, this is not always a comprehensive solution,
as the distributed scheduler greedily assigns root tasks over all workers.
For complex graphs, this can create {\it backpressure} when workers
load excess data, leaving them with insufficient memory
for the results of currently processing tasks.

\begin{figure*}
\begin{center}
\includegraphics*[width=0.75\textwidth]{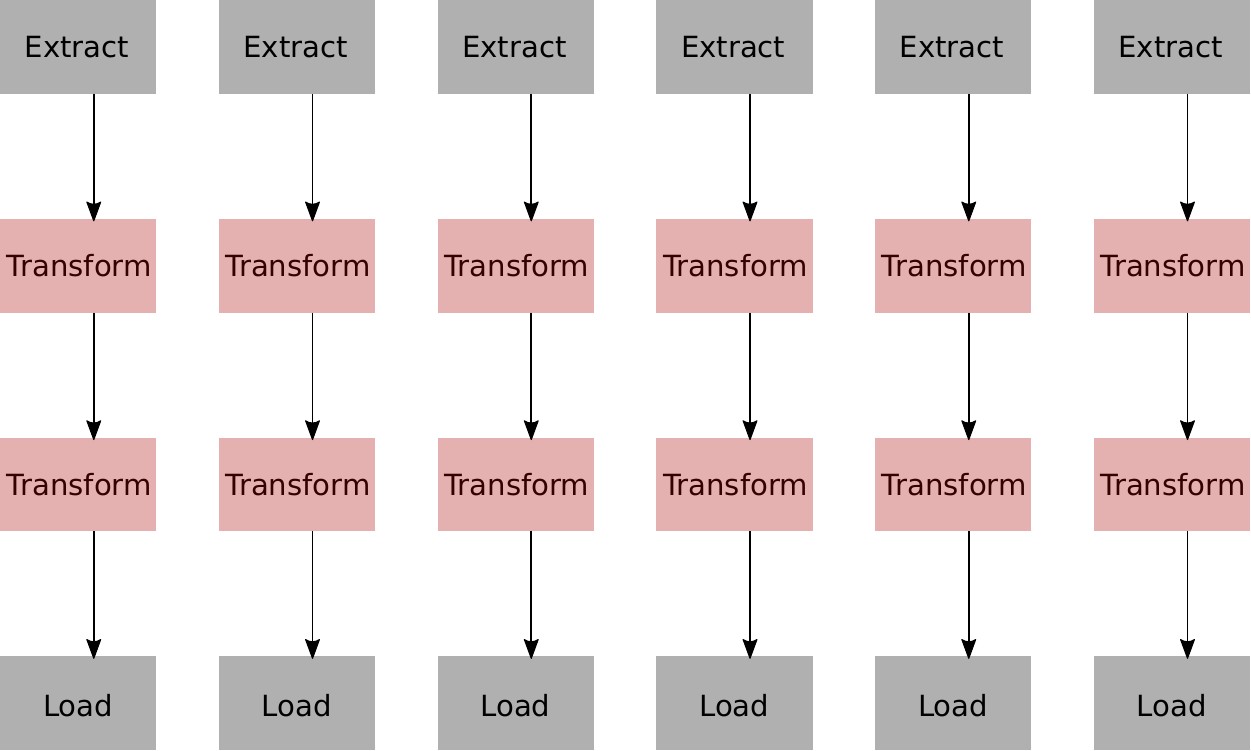}
\end{center}
\caption{Typical Collection Task Flow}
\label{fig:collection_task_flow}
\end{figure*}

{\bf Lack of Pipelining Support for Collections.}
One solution for {\it backpressure} is to use {\it pipelining}
in conjunction with {\it buffering}.
Pipelining involves processing independent chunks of data in stages,
while buffering involves storing a limited number of results
from a previous stage for consumption by later pipeline stages.
Pipelining is implicit in the definition of \dask\ collections:
functions are mapped over chunks of a collection to produce
chunks of a new collection (Figure \ref{fig:collection_task_flow}).
However, buffering of collections is not supported, thus there
is no limit placed on the number of results of a prior stage
that will be computed.

{\bf Chunk Sizes.} Collection chunk sizes need to be sufficiently large,
as this allows the Python interpreter to drop the GIL for a sufficient period of time
for C/C++ Extensions, \numpy\ or \numba\ to fully exercise a CPU core.
The \dask\ documentation recommends they be in the order of several hundred megabytes,
but if the extension utilises multiple cores via the use of a \numba\ {\tt prange} or
\openmp\ loop, then chunk sizes need to be on the order of several gigabytes.
Increasing chunk sizes to both amortise the overhead of a GIL context switch and
fully exercise CPU cores further compounds the effects of backpressure.

{\bf Client Interface.}
While the collection interface is compelling from an HCI perspective,
the fact that collections provide no mechanism for regulating backpressure
makes them ill-suited for more complex distributed processing applications.
Fortunately, \dask\ provides a Client interface where tasks can be submitted
on an executor which returns a distributed future.
This provides the developer with more control over (1) when tasks are submitted
(2) how many tasks are submitted and (3) the order in which tasks execute.

For example, in the case of \pfbimager\ \citep{africanus3}:

\begin{itemize}
    \item Visibility data is sharded over multiple nodes and gridded once.
    \item Deconvolution occurs on a single (large) node with many cores
    and significant amounts of memory.
    \item The deconvolution node needs to
    communicate relatively small amounts of data to the visibility nodes
    between major cycles.
\end{itemize}

Representing the above iterative workflow through a dataflow paradigm
is clumsy, and makes is difficult to ensure that \dask\ respects
memory budgets. Therefore, \pfbimager\ uses stateful \dask\ actors on the abovementioned nodes
that communicate with each other via tasks submitted via the Client interface.

{\bf radio astronomy workflows on AWS EKS.}
We have shown that compute bound algorithms can be made to scale appropriately on AWS in $\S$\ref{section:results_dist_predict}.
However, it is also worth noting that AWS S3 is designed to scale linearly with
the number of nodes and requests.
For single Jones terms, \QuartiCal\ \citep{africanus2} is I/O bound on S3 access,
and we have found that it scales linearly in these cases.

Our experience is that compute on AWS is relatively cheap.
For example, the longest weak-scaling experiment ran for 2975 seconds on 100
{\tt m5.4xlarge} EC2 instances in the {\tt af-south-1} availability zone.
This experiment cost \$84:  $100 \times \$1.1016/ \textrm{hour} \times 2975s/3600$
for compute.

\section{Future Work}

\daskms\ uses {\sc python-casacore}\footnote{\url{https://github.com/casacore/python-casacore}} to interact with CASA Measurement Sets.
While {\sc python-casacore} provides a comprehensive Python interface to the CTDS, it transitively suffers from the CTDS drawbacks
mentioned in $\S$\ref{section:background}.
Additionally, it does not drop the Global Interpreter Lock (GIL), which means that the Python interpreter is blocked by
CTDS I/O.
This can be avoided by exporting the Measurement Set to \zarr\ format, but it would still be convenient for application
developers to interact directly with a native MS.
We have begun work on \arcae\footnote{\url{https://github.com/ratt-ru/arcae}}, a new Python binding to the CTDS  which supports
multi-threaded, GIL-free access to the CTDS with arbitrary indexing along multiple indices.

\section{Conclusions}

We have presented a new framework for developing distributed
radio astronomy applications using software from the \pydata\ ecosystem.
Written in Python, this software should be more accessible to a generation of
scientific software developers and data scientists, compared to software
written in older, more complex languages such as C++.

The quantity of data produced by modern radio interferometers
requires distributed computing by default.
To address this requirement we selected \dask\ as a lightweight,
distributed computing framework.
\dask\ collections such as Arrays and Dataframes allow developers
to rapidly develop both multi-core and distributed dataflow applications.
However, collections can mask the complexities of
distributed computing: more memory is used than is strictly required,
through the lack of buffered pipelining in \dask.
Additionally, as \dask's functional paradigm eschews storing state,
mutating large arrays such as visibility data requires
making a copy further exacerbating memory pressure.

Fortunately, more complicated, stateful applications (in particular imaging)
can be implemented through the use of the Client interface, as well
as the use of \dask\ actors.

Our software has been developed to ingest and write data in the
(row, channel, correlation) Measurement Set v2.0 format.
It is often the case in our \numba\ kernels that we convert
arrays into (time, baseline, channel, correlation)
ordering, as this makes the data easier to reason about
for Calibration and Flagging.
Work has begun by NRAO and the SKAO  on development of a new Measurement Set v4.0
specification which specifies the new, regular ordering in storage.
As this will simplify our software considerably, we aim to
implement this format in the future.

\section*{Acknowledgements}

Funding: OMS's and JSK's research is supported by the South African Research Chairs Initiative of the Department of Science and Technology and National Research Foundation (grant No. 81737). 

The MeerKAT telescope is operated by the South African radio astronomy Observatory, which is a facility of the National Research Foundation, an agency of the Department of Science and Innovation. 





\bibliographystyle{elsarticle-harv}
\bibliography{dask} 

\begin{thebibliography}{76}
\expandafter\ifx\csname natexlab\endcsname\relax\def\natexlab#1{#1}\fi
\providecommand{\url}[1]{\texttt{#1}}
\providecommand{\href}[2]{#2}
\providecommand{\path}[1]{#1}
\providecommand{\DOIprefix}{doi:}
\providecommand{\ArXivprefix}{arXiv:}
\providecommand{\URLprefix}{URL: }
\providecommand{\Pubmedprefix}{pmid:}
\providecommand{\doi}[1]{\href{http://dx.doi.org/#1}{\path{#1}}}
\providecommand{\Pubmed}[1]{\href{pmid:#1}{\path{#1}}}
\providecommand{\bibinfo}[2]{#2}
\ifx\xfnm\relax \def\xfnm[#1]{\unskip,\space#1}\fi
\bibitem[{Abernathey et~al.(2017)Abernathey, Paul, Hamman, Rocklin, Lepore, Tippett, Henderson, Seager, May and Vento}]{Abernathey2017}
\bibinfo{author}{Abernathey, R.}, \bibinfo{author}{Paul, K.}, \bibinfo{author}{Hamman, J.}, \bibinfo{author}{Rocklin, M.}, \bibinfo{author}{Lepore, C.}, \bibinfo{author}{Tippett, M.}, \bibinfo{author}{Henderson, N.}, \bibinfo{author}{Seager, R.}, \bibinfo{author}{May, R.}, \bibinfo{author}{Vento, D.D.}, \bibinfo{year}{2017}.
\newblock \bibinfo{title}{{Pangeo NSF Earthcube Proposal}}.
\newblock \URLprefix \url{https://figshare.com/articles/journal_contribution/Pangeo_NSF_Earthcube_Proposal/5361094}, \DOIprefix\doi{10.6084/m9.figshare.5361094.v1}.
\bibitem[{{Abernathey} et~al.(2018){Abernathey}, {Hamman} and {Miles}}]{Zarr2018}
\bibinfo{author}{{Abernathey}, R.P.}, \bibinfo{author}{{Hamman}, J.}, \bibinfo{author}{{Miles}, A.}, \bibinfo{year}{2018}.
\newblock \bibinfo{title}{{Beyond netCDF: Cloud Native Climate Data with Zarr and XArray}}, in: \bibinfo{booktitle}{AGU Fall Meeting Abstracts}, pp. \bibinfo{pages}{IN33A--06}.
\bibitem[{Alted(2010)}]{Alted2010}
\bibinfo{author}{Alted, F.}, \bibinfo{year}{2010}.
\newblock \bibinfo{title}{Why modern cpus are starving and what can be done about it}.
\newblock \bibinfo{journal}{Computing in Science \& Engineering} \bibinfo{volume}{12}, \bibinfo{pages}{68--71}.
\newblock \DOIprefix\doi{10.1109/MCSE.2010.51}.
\bibitem[{Amdahl(1967)}]{Amdahl1967}
\bibinfo{author}{Amdahl, G.M.}, \bibinfo{year}{1967}.
\newblock \bibinfo{title}{Validity of the single processor approach to achieving large scale computing capabilities}, in: \bibinfo{booktitle}{Proceedings of the April 18-20, 1967, Spring Joint Computer Conference}, \bibinfo{publisher}{Association for Computing Machinery}, \bibinfo{address}{New York, NY, USA}. p. \bibinfo{pages}{483–485}.
\newblock \URLprefix \url{https://doi.org/10.1145/1465482.1465560}, \DOIprefix\doi{10.1145/1465482.1465560}.
\bibitem[{Anaconda(2022)}]{Datashader2022}
\bibinfo{author}{Anaconda}, \bibinfo{year}{2022}.
\newblock \bibinfo{title}{Datashader}.
\newblock \bibinfo{howpublished}{\url{https://datashader.org/}}.
\newblock \bibinfo{note}{Online; accessed 14-11-2022}.
\bibitem[{{Andati} et~al.(2022){Andati}, {Smirnov} and {Makhathini}}]{Ragavi2022}
\bibinfo{author}{{Andati}, L.A.L.}, \bibinfo{author}{{Smirnov}, O.M.}, \bibinfo{author}{{Makhathini}, S.}, \bibinfo{year}{2022}.
\newblock \bibinfo{title}{{RAGaVI: A Radio Astronomy Gains and Visibilities Inspector}}, in: \bibinfo{editor}{{Ruiz}, J.E.}, \bibinfo{editor}{{Pierfedereci}, F.}, \bibinfo{editor}{{Teuben}, P.} (Eds.), \bibinfo{booktitle}{Astronomical Society of the Pacific Conference Series}, p. \bibinfo{pages}{529}.
\bibitem[{{Apache Software Foundation.}(2019)}]{Arrow2019}
\bibinfo{author}{{Apache Software Foundation.}}, \bibinfo{year}{2019}.
\newblock \bibinfo{title}{Arrow: a cross-language development platform for in-memory data}.
\newblock \URLprefix \url{https://arrow.apache.org}.
\bibitem[{{Arras, Philipp} et~al.(2021){Arras, Philipp}, {Reinecke, Martin}, {Westermann, Rüdiger} and {Enßlin, Torsten A.}}]{Arras2021}
\bibinfo{author}{{Arras, Philipp}}, \bibinfo{author}{{Reinecke, Martin}}, \bibinfo{author}{{Westermann, Rüdiger}}, \bibinfo{author}{{Enßlin, Torsten A.}}, \bibinfo{year}{2021}.
\newblock \bibinfo{title}{Efficient wide-field radio interferometry response}.
\newblock \bibinfo{journal}{Astronomy and Astrophysics} \bibinfo{volume}{646}, \bibinfo{pages}{A58}.
\newblock \DOIprefix\doi{10.1051/0004-6361/202039723}.
\bibitem[{{Astropy Collaboration} et~al.(2022){Astropy Collaboration}, {Price-Whelan}, {Lim}, {Earl}, {Starkman}, {Bradley}, {Shupe}, {Patil}, {Corrales}, {Brasseur}, {N{\"o}the}, {Donath}, {Tollerud}, {Morris}, {Ginsburg}, {Vaher}, {Weaver}, {Tocknell}, {Jamieson}, {van Kerkwijk}, {Robitaille}, {Merry}, {Bachetti}, {G{\"u}nther}, {Aldcroft}, {Alvarado-Montes}, {Archibald}, {B{\'o}di}, {Bapat}, {Barentsen}, {Baz{\'a}n}, {Biswas}, {Boquien}, {Burke}, {Cara}, {Cara}, {Conroy}, {Conseil}, {Craig}, {Cross}, {Cruz}, {D'Eugenio}, {Dencheva}, {Devillepoix}, {Dietrich}, {Eigenbrot}, {Erben}, {Ferreira}, {Foreman-Mackey}, {Fox}, {Freij}, {Garg}, {Geda}, {Glattly}, {Gondhalekar}, {Gordon}, {Grant}, {Greenfield}, {Groener}, {Guest}, {Gurovich}, {Handberg}, {Hart}, {Hatfield-Dodds}, {Homeier}, {Hosseinzadeh}, {Jenness}, {Jones}, {Joseph}, {Kalmbach}, {Karamehmetoglu}, {Ka{\l}uszy{\'n}ski}, {Kelley}, {Kern}, {Kerzendorf}, {Koch}, {Kulumani}, {Lee}, {Ly}, {Ma}, {MacBride}, {Maljaars}, {Muna}, {Murphy}, {Norman}, {O'Steen}, {Oman}, {Pacifici}, {Pascual}, {Pascual-Granado}, {Patil}, {Perren}, {Pickering}, {Rastogi}, {Roulston}, {Ryan}, {Rykoff}, {Sabater}, {Sakurikar}, {Salgado}, {Sanghi}, {Saunders}, {Savchenko}, {Schwardt}, {Seifert-Eckert}, {Shih}, {Jain}, {Shukla}, {Sick}, {Simpson}, {Singanamalla}, {Singer}, {Singhal}, {Sinha}, {Sip{\H{o}}cz}, {Spitler}, {Stansby}, {Streicher}, {{\v{S}}umak}, {Swinbank}, {Taranu}, {Tewary}, {Tremblay}, {de Val-Borro}, {Van Kooten}, {Vasovi{\'c}}, {Verma}, {de Miranda Cardoso}, {Williams}, {Wilson}, {Winkel}, {Wood-Vasey}, {Xue}, {Yoachim}, {Zhang}, {Zonca} and {Astropy Project Contributors}}]{Astropy2022}
\bibinfo{author}{{Astropy Collaboration}}, \bibinfo{author}{{Price-Whelan}, A.M.}, \bibinfo{author}{{Lim}, P.L.}, \bibinfo{author}{{Earl}, N.}, \bibinfo{author}{{Starkman}, N.}, \bibinfo{author}{{Bradley}, L.}, \bibinfo{author}{{Shupe}, D.L.}, \bibinfo{author}{{Patil}, A.A.}, \bibinfo{author}{{Corrales}, L.}, \bibinfo{author}{{Brasseur}, C.E.}, \bibinfo{author}{{N{\"o}the}, M.}, \bibinfo{author}{{Donath}, A.}, \bibinfo{author}{{Tollerud}, E.}, \bibinfo{author}{{Morris}, B.M.}, \bibinfo{author}{{Ginsburg}, A.}, \bibinfo{author}{{Vaher}, E.}, \bibinfo{author}{{Weaver}, B.A.}, \bibinfo{author}{{Tocknell}, J.}, \bibinfo{author}{{Jamieson}, W.}, \bibinfo{author}{{van Kerkwijk}, M.H.}, \bibinfo{author}{{Robitaille}, T.P.}, \bibinfo{author}{{Merry}, B.}, \bibinfo{author}{{Bachetti}, M.}, \bibinfo{author}{{G{\"u}nther}, H.M.}, \bibinfo{author}{{Aldcroft}, T.L.}, \bibinfo{author}{{Alvarado-Montes}, J.A.}, \bibinfo{author}{{Archibald}, A.M.}, \bibinfo{author}{{B{\'o}di}, A.}, \bibinfo{author}{{Bapat}, S.}, \bibinfo{author}{{Barentsen}, G.}, \bibinfo{author}{{Baz{\'a}n}, J.}, \bibinfo{author}{{Biswas}, M.}, \bibinfo{author}{{Boquien}, M.}, \bibinfo{author}{{Burke}, D.J.}, \bibinfo{author}{{Cara}, D.}, \bibinfo{author}{{Cara}, M.}, \bibinfo{author}{{Conroy}, K.E.}, \bibinfo{author}{{Conseil}, S.}, \bibinfo{author}{{Craig}, M.W.}, \bibinfo{author}{{Cross}, R.M.}, \bibinfo{author}{{Cruz}, K.L.}, \bibinfo{author}{{D'Eugenio}, F.}, \bibinfo{author}{{Dencheva}, N.}, \bibinfo{author}{{Devillepoix}, H.A.R.}, \bibinfo{author}{{Dietrich}, J.P.}, \bibinfo{author}{{Eigenbrot}, A.D.}, \bibinfo{author}{{Erben}, T.}, \bibinfo{author}{{Ferreira}, L.}, \bibinfo{author}{{Foreman-Mackey}, D.}, \bibinfo{author}{{Fox}, R.}, \bibinfo{author}{{Freij}, N.}, \bibinfo{author}{{Garg}, S.}, \bibinfo{author}{{Geda}, R.}, \bibinfo{author}{{Glattly}, L.}, \bibinfo{author}{{Gondhalekar}, Y.}, \bibinfo{author}{{Gordon}, K.D.}, \bibinfo{author}{{Grant}, D.}, \bibinfo{author}{{Greenfield}, P.}, \bibinfo{author}{{Groener}, A.M.}, \bibinfo{author}{{Guest}, S.}, \bibinfo{author}{{Gurovich}, S.}, \bibinfo{author}{{Handberg}, R.}, \bibinfo{author}{{Hart}, A.}, \bibinfo{author}{{Hatfield-Dodds}, Z.}, \bibinfo{author}{{Homeier}, D.}, \bibinfo{author}{{Hosseinzadeh}, G.}, \bibinfo{author}{{Jenness}, T.}, \bibinfo{author}{{Jones}, C.K.}, \bibinfo{author}{{Joseph}, P.}, \bibinfo{author}{{Kalmbach}, J.B.}, \bibinfo{author}{{Karamehmetoglu}, E.}, \bibinfo{author}{{Ka{\l}uszy{\'n}ski}, M.}, \bibinfo{author}{{Kelley}, M.S.P.}, \bibinfo{author}{{Kern}, N.}, \bibinfo{author}{{Kerzendorf}, W.E.}, \bibinfo{author}{{Koch}, E.W.}, \bibinfo{author}{{Kulumani}, S.}, \bibinfo{author}{{Lee}, A.}, \bibinfo{author}{{Ly}, C.}, \bibinfo{author}{{Ma}, Z.}, \bibinfo{author}{{MacBride}, C.}, \bibinfo{author}{{Maljaars}, J.M.}, \bibinfo{author}{{Muna}, D.}, \bibinfo{author}{{Murphy}, N.A.}, \bibinfo{author}{{Norman}, H.}, \bibinfo{author}{{O'Steen}, R.}, \bibinfo{author}{{Oman}, K.A.}, \bibinfo{author}{{Pacifici}, C.}, \bibinfo{author}{{Pascual}, S.}, \bibinfo{author}{{Pascual-Granado}, J.}, \bibinfo{author}{{Patil}, R.R.}, \bibinfo{author}{{Perren}, G.I.}, \bibinfo{author}{{Pickering}, T.E.}, \bibinfo{author}{{Rastogi}, T.}, \bibinfo{author}{{Roulston}, B.R.}, \bibinfo{author}{{Ryan}, D.F.}, \bibinfo{author}{{Rykoff}, E.S.}, \bibinfo{author}{{Sabater}, J.}, \bibinfo{author}{{Sakurikar}, P.}, \bibinfo{author}{{Salgado}, J.}, \bibinfo{author}{{Sanghi}, A.}, \bibinfo{author}{{Saunders}, N.}, \bibinfo{author}{{Savchenko}, V.}, \bibinfo{author}{{Schwardt}, L.}, \bibinfo{author}{{Seifert-Eckert}, M.}, \bibinfo{author}{{Shih}, A.Y.}, \bibinfo{author}{{Jain}, A.S.}, \bibinfo{author}{{Shukla}, G.}, \bibinfo{author}{{Sick}, J.}, \bibinfo{author}{{Simpson}, C.}, \bibinfo{author}{{Singanamalla}, S.}, \bibinfo{author}{{Singer}, L.P.}, \bibinfo{author}{{Singhal}, J.}, \bibinfo{author}{{Sinha}, M.}, \bibinfo{author}{{Sip{\H{o}}cz}, B.M.}, \bibinfo{author}{{Spitler}, L.R.}, \bibinfo{author}{{Stansby}, D.}, \bibinfo{author}{{Streicher}, O.}, \bibinfo{author}{{{\v{S}}umak}, J.}, \bibinfo{author}{{Swinbank}, J.D.}, \bibinfo{author}{{Taranu}, D.S.}, \bibinfo{author}{{Tewary}, N.}, \bibinfo{author}{{Tremblay}, G.R.}, \bibinfo{author}{{de Val-Borro}, M.}, \bibinfo{author}{{Van Kooten}, S.J.}, \bibinfo{author}{{Vasovi{\'c}}, Z.}, \bibinfo{author}{{Verma}, S.}, \bibinfo{author}{{de Miranda Cardoso}, J.V.}, \bibinfo{author}{{Williams}, P.K.G.}, \bibinfo{author}{{Wilson}, T.J.}, \bibinfo{author}{{Winkel}, B.}, \bibinfo{author}{{Wood-Vasey}, W.M.}, \bibinfo{author}{{Xue}, R.}, \bibinfo{author}{{Yoachim}, P.}, \bibinfo{author}{{Zhang}, C.}, \bibinfo{author}{{Zonca}, A.}, \bibinfo{author}{{Astropy Project Contributors}}, \bibinfo{year}{2022}.
\newblock \bibinfo{title}{{The Astropy Project: Sustaining and Growing a Community-oriented Open-source Project and the Latest Major Release (v5.0) of the Core Package}}.
\newblock \bibinfo{journal}{The Astrophysical Journal} \bibinfo{volume}{935}, \bibinfo{pages}{167}.
\newblock \DOIprefix\doi{10.3847/1538-4357/ac7c74}, \href{http://arxiv.org/abs/2206.14220}{{\tt arXiv:2206.14220}}.
\bibitem[{Atemkeng et~al.(2022)Atemkeng, Perkins, Kenyon, Hugo and Smirnov}]{atemkeng2021}
\bibinfo{author}{Atemkeng, M.}, \bibinfo{author}{Perkins, S.}, \bibinfo{author}{Kenyon, J.}, \bibinfo{author}{Hugo, B.}, \bibinfo{author}{Smirnov, O.M.}, \bibinfo{year}{2022}.
\newblock \bibinfo{title}{Xova: Baseline-dependent time and channel averaging for radio interferometry}, in: \bibinfo{editor}{{Ruiz}, J.E.}, \bibinfo{editor}{{Pierfedereci}, F.}, \bibinfo{editor}{{Teuben}, P.} (Eds.), \bibinfo{booktitle}{Astronomical Society of the Pacific Conference Series}.
\bibitem[{{Atemkeng} et~al.(2016){Atemkeng}, {Smirnov}, {Tasse}, {Foster} and {Jonas}}]{Atemkeng2016}
\bibinfo{author}{{Atemkeng}, M.T.}, \bibinfo{author}{{Smirnov}, O.M.}, \bibinfo{author}{{Tasse}, C.}, \bibinfo{author}{{Foster}, G.}, \bibinfo{author}{{Jonas}, J.}, \bibinfo{year}{2016}.
\newblock \bibinfo{title}{{Using baseline-dependent window functions for data compression and field-of-interest shaping in radio interferometry}}.
\newblock \bibinfo{journal}{Monthly Notices of the Royal Astronomical Society} \bibinfo{volume}{462}, \bibinfo{pages}{2542--2558}.
\newblock \DOIprefix\doi{10.1093/mnras/stw1656}, \href{http://arxiv.org/abs/1607.04106}{{\tt arXiv:1607.04106}}.
\bibitem[{Bester et~al.(2024)Bester, Kenyon, Repetti, Perkins, Smirnov, Blecher, Mhiri, Roth, Heywood, Wiaux and Hugo}]{africanus3}
\bibinfo{author}{Bester, H.L.}, \bibinfo{author}{Kenyon, J.S.}, \bibinfo{author}{Repetti, A.}, \bibinfo{author}{Perkins, S.J.}, \bibinfo{author}{Smirnov, O.M.}, \bibinfo{author}{Blecher, T.}, \bibinfo{author}{Mhiri, Y.}, \bibinfo{author}{Roth, J.}, \bibinfo{author}{Heywood, I.}, \bibinfo{author}{Wiaux, Y.}, \bibinfo{author}{Hugo, B.V.}, \bibinfo{year}{2024}.
\newblock \bibinfo{title}{{Africanus III. pfb-imaging -- a flexible radio interferometric imaging suite}}.
\newblock \bibinfo{journal}{Astronomy and Computing} \bibinfo{volume}{submitted}.
\newblock \href{http://arxiv.org/abs/2412.10073}{{\tt arXiv:2412.10073}}.
\bibitem[{Bezanson et~al.(2017)Bezanson, Edelman, Karpinski and Shah}]{Julia-2017}
\bibinfo{author}{Bezanson, J.}, \bibinfo{author}{Edelman, A.}, \bibinfo{author}{Karpinski, S.}, \bibinfo{author}{Shah, V.B.}, \bibinfo{year}{2017}.
\newblock \bibinfo{title}{Julia: A fresh approach to numerical computing}.
\newblock \bibinfo{journal}{SIAM {R}eview} \bibinfo{volume}{59}, \bibinfo{pages}{65--98}.
\newblock \DOIprefix\doi{10.1137/141000671}.
\bibitem[{Bradbury et~al.(2018)Bradbury, Frostig, Hawkins, Johnson, Leary, Maclaurin, Necula, Paszke, Vander{P}las, Wanderman-{M}ilne and Zhang}]{jax2018}
\bibinfo{author}{Bradbury, J.}, \bibinfo{author}{Frostig, R.}, \bibinfo{author}{Hawkins, P.}, \bibinfo{author}{Johnson, M.J.}, \bibinfo{author}{Leary, C.}, \bibinfo{author}{Maclaurin, D.}, \bibinfo{author}{Necula, G.}, \bibinfo{author}{Paszke, A.}, \bibinfo{author}{Vander{P}las, J.}, \bibinfo{author}{Wanderman-{M}ilne, S.}, \bibinfo{author}{Zhang, Q.}, \bibinfo{year}{2018}.
\newblock \bibinfo{title}{{JAX}: composable transformations of {P}ython+{N}um{P}y programs}.
\newblock \URLprefix \url{http://github.com/google/jax}.
\bibitem[{Byrne and Jacobs(2021)}]{cloud21cmcosmology2021}
\bibinfo{author}{Byrne, R.}, \bibinfo{author}{Jacobs, D.}, \bibinfo{year}{2021}.
\newblock \bibinfo{title}{Development of a high throughput cloud-based data pipeline for 21 cm cosmology}.
\newblock \bibinfo{journal}{Astronomy and Computing} \bibinfo{volume}{34}, \bibinfo{pages}{100447}.
\newblock \DOIprefix\doi{https://doi.org/10.1016/j.ascom.2021.100447}.
\bibitem[{Chy{\.{z} }y et~al.(2018)Chy{\.{z} }y, Jurusik, Piotrowska, Nikiel-Wroczy{\'{n}}ski, Heesen, Vacca, Nowak, Paladino, Surma, Sridhar, Heald, Beck, Conway, Sendlinger, Cury{\l}o, Mulcahy, Broderick, Hardcastle, Callingham, Gürkan, Iacobelli, Röttgering, Adebahr, Shulevski, Dettmar, Breton, Clarke, Farnes, Orr{\'{u}}, Pandey, Pandey-Pommier, Pizzo, Riseley, Rowlinson, Scaife, Stewart, van~der Horst and van Weeren}]{msss2018}
\bibinfo{author}{Chy{\.{z} }y, K.T.}, \bibinfo{author}{Jurusik, W.}, \bibinfo{author}{Piotrowska, J.}, \bibinfo{author}{Nikiel-Wroczy{\'{n}}ski, B.}, \bibinfo{author}{Heesen, V.}, \bibinfo{author}{Vacca, V.}, \bibinfo{author}{Nowak, N.}, \bibinfo{author}{Paladino, R.}, \bibinfo{author}{Surma, P.}, \bibinfo{author}{Sridhar, S.S.}, \bibinfo{author}{Heald, G.}, \bibinfo{author}{Beck, R.}, \bibinfo{author}{Conway, J.}, \bibinfo{author}{Sendlinger, K.}, \bibinfo{author}{Cury{\l}o, M.}, \bibinfo{author}{Mulcahy, D.}, \bibinfo{author}{Broderick, J.W.}, \bibinfo{author}{Hardcastle, M.J.}, \bibinfo{author}{Callingham, J.R.}, \bibinfo{author}{Gürkan, G.}, \bibinfo{author}{Iacobelli, M.}, \bibinfo{author}{Röttgering, H.J.A.}, \bibinfo{author}{Adebahr, B.}, \bibinfo{author}{Shulevski, A.}, \bibinfo{author}{Dettmar, R.J.}, \bibinfo{author}{Breton, R.P.}, \bibinfo{author}{Clarke, A.O.}, \bibinfo{author}{Farnes, J.S.}, \bibinfo{author}{Orr{\'{u}}, E.}, \bibinfo{author}{Pandey, V.N.}, \bibinfo{author}{Pandey-Pommier, M.}, \bibinfo{author}{Pizzo, R.}, \bibinfo{author}{Riseley, C.J.}, \bibinfo{author}{Rowlinson, A.}, \bibinfo{author}{Scaife, A.M.M.}, \bibinfo{author}{Stewart, A.J.}, \bibinfo{author}{van~der Horst, A.J.}, \bibinfo{author}{van Weeren, R.J.}, \bibinfo{year}{2018}.
\newblock \bibinfo{title}{{LOFAR} {MSSS}: Flattening low-frequency radio continuum spectra of nearby galaxies}.
\newblock \bibinfo{journal}{Astronomy and Astrophysics} \bibinfo{volume}{619}, \bibinfo{pages}{A36}.
\newblock \DOIprefix\doi{10.1051/0004-6361/201833133}.
\bibitem[{{Cornwell} and {Perley}(1992)}]{Cornwell1992}
\bibinfo{author}{{Cornwell}, T.J.}, \bibinfo{author}{{Perley}, R.A.}, \bibinfo{year}{1992}.
\newblock \bibinfo{title}{{Radio-interferometric imaging of very large fields. The problem of non-coplanar arrays.}}
\newblock \bibinfo{journal}{Astronomy and Astrophysics} \bibinfo{volume}{261}, \bibinfo{pages}{353--364}.
\bibitem[{{Dask Development Team}(2016)}]{Dask2016}
\bibinfo{author}{{Dask Development Team}}, \bibinfo{year}{2016}.
\newblock \bibinfo{title}{Dask: Library for dynamic task scheduling}.
\newblock \URLprefix \url{https://dask.org}.
\bibitem[{{Dask Development Team}(2023)}]{DaskDistributedScheduling1999}
\bibinfo{author}{{Dask Development Team}}, \bibinfo{year}{2023}.
\newblock \bibinfo{title}{Scheduling policies}.
\newblock \bibinfo{howpublished}{\url{https://web.archive.org/web/20230928002741/https://distributed.dask.org/en/stable/scheduling-policies.html}}.
\newblock \bibinfo{note}{Accessed: 2023-09-28}.
\bibitem[{{Dask Development Team}(2024)}]{DaskLayerAnnotations2020}
\bibinfo{author}{{Dask Development Team}}, \bibinfo{year}{2024}.
\newblock \bibinfo{title}{Dask layer annotations}.
\newblock \bibinfo{howpublished}{\url{https://github.com/dask/dask/pull/6767}}.
\newblock \bibinfo{note}{Accessed: 2024-02-07}.
\bibitem[{Dennis(1974)}]{Dennis1974a}
\bibinfo{author}{Dennis, J.B.}, \bibinfo{year}{1974}.
\newblock \bibinfo{title}{First version of a data flow procedure language}, in: \bibinfo{booktitle}{Programming Symposium, Proceedings Colloque Sur La Programmation}, \bibinfo{publisher}{Springer-Verlag}, \bibinfo{address}{Berlin, Heidelberg}. p. \bibinfo{pages}{362–376}.
\bibitem[{Dennis and Misunas(1974)}]{Dennis1974b}
\bibinfo{author}{Dennis, J.B.}, \bibinfo{author}{Misunas, D.P.}, \bibinfo{year}{1974}.
\newblock \bibinfo{title}{A preliminary architecture for a basic data-flow processor}, in: \bibinfo{booktitle}{Proceedings of the 2nd Annual Symposium on Computer Architecture}, \bibinfo{publisher}{Association for Computing Machinery}, \bibinfo{address}{New York, NY, USA}. p. \bibinfo{pages}{126–132}.
\newblock \DOIprefix\doi{10.1145/642089.642111}.
\bibitem[{{Di Francesco} et~al.(2019){Di Francesco}, {Chalmers}, {Denman}, {Fissel}, {Friesen}, {Gaensler}, {Hlavacek-Larrondo}, {Kirk}, {Matthews}, {O'Dea}, {Robishaw}, {Rosolowsky}, {Rupen}, {Sadavoy}, {Sa-Harb}, {Sivakoff}, {Tahani}, {van der Marel}, {White} and {Wilson}}]{ngvla2019}
\bibinfo{author}{{Di Francesco}, J.}, \bibinfo{author}{{Chalmers}, D.}, \bibinfo{author}{{Denman}, N.}, \bibinfo{author}{{Fissel}, L.}, \bibinfo{author}{{Friesen}, R.}, \bibinfo{author}{{Gaensler}, B.}, \bibinfo{author}{{Hlavacek-Larrondo}, J.}, \bibinfo{author}{{Kirk}, H.}, \bibinfo{author}{{Matthews}, B.}, \bibinfo{author}{{O'Dea}, C.}, \bibinfo{author}{{Robishaw}, T.}, \bibinfo{author}{{Rosolowsky}, E.}, \bibinfo{author}{{Rupen}, M.}, \bibinfo{author}{{Sadavoy}, S.}, \bibinfo{author}{{Sa-Harb}, S.}, \bibinfo{author}{{Sivakoff}, G.}, \bibinfo{author}{{Tahani}, M.}, \bibinfo{author}{{van der Marel}, N.}, \bibinfo{author}{{White}, J.}, \bibinfo{author}{{Wilson}, C.}, \bibinfo{year}{2019}.
\newblock \bibinfo{title}{{The Next Generation Very Large Array}}, in: \bibinfo{booktitle}{Canadian Long Range Plan for Astronomy and Astrophysics White Papers}, p.~\bibinfo{pages}{32}.
\newblock \DOIprefix\doi{10.5281/zenodo.3765763}, \href{http://arxiv.org/abs/1911.01517}{{\tt arXiv:1911.01517}}.
\bibitem[{van Diepen(2015)}]{Diepen2015}
\bibinfo{author}{van Diepen, G.}, \bibinfo{year}{2015}.
\newblock \bibinfo{title}{Casacore table data system and its use in the measurementset}.
\newblock \bibinfo{journal}{Astronomy and Computing} \bibinfo{volume}{2}.
\newblock \DOIprefix\doi{10.1016/j.ascom.2015.06.002}.
\bibitem[{Diepen(2006)}]{VanDiepen2006}
\bibinfo{author}{Diepen, G.V.}, \bibinfo{year}{2006}.
\newblock \bibinfo{title}{{NOTE 199 - Table Query Language}}.
\newblock \bibinfo{type}{Technical Report}. AIPS++ Notes Series.
\bibitem[{{Dodson} et~al.(2022){Dodson}, {Momjian}, {Pisano}, {Luber}, {Blue Bird}, {Rozgonyi}, {Smith}, {van Gorkom}, {Lucero}, {Hess}, {Yun}, {Rhee}, {van der Hulst}, {Vinsen}, {Meyer}, {Fernandez}, {Gim}, {Popping} and {Wilcots}}]{dodson-cloud2}
\bibinfo{author}{{Dodson}, R.}, \bibinfo{author}{{Momjian}, E.}, \bibinfo{author}{{Pisano}, D.J.}, \bibinfo{author}{{Luber}, N.}, \bibinfo{author}{{Blue Bird}, J.}, \bibinfo{author}{{Rozgonyi}, K.}, \bibinfo{author}{{Smith}, E.T.}, \bibinfo{author}{{van Gorkom}, J.H.}, \bibinfo{author}{{Lucero}, D.}, \bibinfo{author}{{Hess}, K.M.}, \bibinfo{author}{{Yun}, M.}, \bibinfo{author}{{Rhee}, J.}, \bibinfo{author}{{van der Hulst}, J.M.}, \bibinfo{author}{{Vinsen}, K.}, \bibinfo{author}{{Meyer}, M.}, \bibinfo{author}{{Fernandez}, X.}, \bibinfo{author}{{Gim}, H.B.}, \bibinfo{author}{{Popping}, A.}, \bibinfo{author}{{Wilcots}, E.}, \bibinfo{year}{2022}.
\newblock \bibinfo{title}{{CHILES. VII. Deep Imaging for the CHILES Project, an SKA Prototype}}.
\newblock \bibinfo{journal}{Astrophysics Journal} \bibinfo{volume}{163}, \bibinfo{pages}{59}.
\newblock \DOIprefix\doi{10.3847/1538-3881/ac3e65}, \href{http://arxiv.org/abs/2112.06488}{{\tt arXiv:2112.06488}}.
\bibitem[{{Dodson} et~al.(2016){Dodson}, {Vinsen}, {Wu}, {Popping}, {Meyer}, {Wicenec}, {Quinn}, {van Gorkom} and {Momjian}}]{dodson-cloud}
\bibinfo{author}{{Dodson}, R.}, \bibinfo{author}{{Vinsen}, K.}, \bibinfo{author}{{Wu}, C.}, \bibinfo{author}{{Popping}, A.}, \bibinfo{author}{{Meyer}, M.}, \bibinfo{author}{{Wicenec}, A.}, \bibinfo{author}{{Quinn}, P.}, \bibinfo{author}{{van Gorkom}, J.}, \bibinfo{author}{{Momjian}, E.}, \bibinfo{year}{2016}.
\newblock \bibinfo{title}{{Imaging SKA-scale data in three different computing environments}}.
\newblock \bibinfo{journal}{Astronomy and Computing} \bibinfo{volume}{14}, \bibinfo{pages}{8--22}.
\newblock \DOIprefix\doi{10.1016/j.ascom.2015.10.007}, \href{http://arxiv.org/abs/1511.00401}{{\tt arXiv:1511.00401}}.
\bibitem[{{Finlay} et~al.(2023){Finlay}, {Bassett}, {Kunz} and {Oozeer}}]{Finlay2023}
\bibinfo{author}{{Finlay}, C.}, \bibinfo{author}{{Bassett}, B.A.}, \bibinfo{author}{{Kunz}, M.}, \bibinfo{author}{{Oozeer}, N.}, \bibinfo{year}{2023}.
\newblock \bibinfo{title}{{Trajectory-based RFI subtraction and calibration for radio interferometry}}.
\newblock \bibinfo{journal}{Monthly Notices of the Royal Astronomical Society} \bibinfo{volume}{524}, \bibinfo{pages}{3231--3251}.
\newblock \DOIprefix\doi{10.1093/mnras/stad1979}, \href{http://arxiv.org/abs/2301.04188}{{\tt arXiv:2301.04188}}.
\bibitem[{{Greisen}(1990)}]{aips1990}
\bibinfo{author}{{Greisen}, E.W.}, \bibinfo{year}{1990}.
\newblock \bibinfo{title}{{The Astronomical Image Processing System.}}, in: \bibinfo{booktitle}{Acquisition, Processing and Archiving of Astronomical Images}, pp. \bibinfo{pages}{125--142}.
\bibitem[{Gustafson(1990)}]{Gustafson1990}
\bibinfo{author}{Gustafson, J.L.}, \bibinfo{year}{1990}.
\newblock \bibinfo{title}{Fixed time, tiered memory, and superlinear speedup}, in: \bibinfo{booktitle}{Proceedings of the Fifth Distributed Memory Computing Conference (DMCC5)}, \bibinfo{organization}{IEEE Press}. pp. \bibinfo{pages}{1255--1260}.
\bibitem[{van Haarlem et~al.(2013)van Haarlem, Wise, Gunst, Heald, McKean, Hessels, de~Bruyn, Nijboer, Swinbank, Fallows, Brentjens, Nelles, Beck, Falcke, Fender, Hörandel, Koopmans, Mann, Miley, Röttgering, Stappers, Wijers, Zaroubi, van~den Akker, Alexov, Anderson, Anderson, van Ardenne, Arts, Asgekar, Avruch, Batejat, Bähren, Bell, Bell, van Bemmel, Bennema, Bentum, Bernardi, Best, B{\^{\i} }rzan, Bonafede, Boonstra, Braun, Bregman, Breitling, van~de Brink, Broderick, Broekema, Brouw, Brüggen, Butcher, van Cappellen, Ciardi, Coenen, Conway, Coolen, Corstanje, Damstra, Davies, Deller, Dettmar, van Diepen, Dijkstra, Donker, Doorduin, Dromer, Drost, van Duin, Eislöffel, van Enst, Ferrari, Frieswijk, Gankema, Garrett, de~Gasperin, Gerbers, de~Geus, Grie{\ss}meier, Grit, Gruppen, Hamaker, Hassall, Hoeft, Holties, Horneffer, van~der Horst, van Houwelingen, Huijgen, Iacobelli, Intema, Jackson, Jelic, de~Jong, Juette, Kant, Karastergiou, Koers, Kollen, Kondratiev, Kooistra, Koopman, Koster, Kuniyoshi, Kramer, Kuper, Lambropoulos, Law, van Leeuwen, Lemaitre, Loose, Maat, Macario, Markoff, Masters, McFadden, McKay-Bukowski, Meijering, Meulman, Mevius, Middelberg, Millenaar, Miller-Jones, Mohan, Mol, Morawietz, Morganti, Mulcahy, Mulder, Munk, Nieuwenhuis, van Nieuwpoort, Noordam, Norden, Noutsos, Offringa, Olofsson, Omar, Orr{\'{u}}, Overeem, Paas, Pandey-Pommier, Pandey, Pizzo, Polatidis, Rafferty, Rawlings, Reich, de~Reijer, Reitsma, Renting, Riemers, Rol, Romein, Roosjen, Ruiter, Scaife, van~der Schaaf, Scheers, Schellart, Schoenmakers, Schoonderbeek, Serylak, Shulevski, Sluman, Smirnov, Sobey, Spreeuw, Steinmetz, Sterks, Stiepel, Stuurwold, Tagger, Tang, Tasse, Thomas, Thoudam, Toribio, van~der Tol, Usov, van Veelen, van~der Veen, ter Veen, Verbiest, Vermeulen, Vermaas, Vocks, Vogt, de~Vos, van~der Wal, van Weeren, Weggemans, Weltevrede, White, Wijnholds, Wilhelmsson, Wucknitz, Yatawatta, Zarka, Zensus and van Zwieten}]{van_Haarlem_2013}
\bibinfo{author}{van Haarlem, M.P.}, \bibinfo{author}{Wise, M.W.}, \bibinfo{author}{Gunst, A.W.}, \bibinfo{author}{Heald, G.}, \bibinfo{author}{McKean, J.P.}, \bibinfo{author}{Hessels, J.W.T.}, \bibinfo{author}{de~Bruyn, A.G.}, \bibinfo{author}{Nijboer, R.}, \bibinfo{author}{Swinbank, J.}, \bibinfo{author}{Fallows, R.}, \bibinfo{author}{Brentjens, M.}, \bibinfo{author}{Nelles, A.}, \bibinfo{author}{Beck, R.}, \bibinfo{author}{Falcke, H.}, \bibinfo{author}{Fender, R.}, \bibinfo{author}{Hörandel, J.}, \bibinfo{author}{Koopmans, L.V.E.}, \bibinfo{author}{Mann, G.}, \bibinfo{author}{Miley, G.}, \bibinfo{author}{Röttgering, H.}, \bibinfo{author}{Stappers, B.W.}, \bibinfo{author}{Wijers, R.A.M.J.}, \bibinfo{author}{Zaroubi, S.}, \bibinfo{author}{van~den Akker, M.}, \bibinfo{author}{Alexov, A.}, \bibinfo{author}{Anderson, J.}, \bibinfo{author}{Anderson, K.}, \bibinfo{author}{van Ardenne, A.}, \bibinfo{author}{Arts, M.}, \bibinfo{author}{Asgekar, A.}, \bibinfo{author}{Avruch, I.M.}, \bibinfo{author}{Batejat, F.}, \bibinfo{author}{Bähren, L.}, \bibinfo{author}{Bell, M.E.}, \bibinfo{author}{Bell, M.R.}, \bibinfo{author}{van Bemmel, I.}, \bibinfo{author}{Bennema, P.}, \bibinfo{author}{Bentum, M.J.}, \bibinfo{author}{Bernardi, G.}, \bibinfo{author}{Best, P.}, \bibinfo{author}{B{\^{\i} }rzan, L.}, \bibinfo{author}{Bonafede, A.}, \bibinfo{author}{Boonstra, A.J.}, \bibinfo{author}{Braun, R.}, \bibinfo{author}{Bregman, J.}, \bibinfo{author}{Breitling, F.}, \bibinfo{author}{van~de Brink, R.H.}, \bibinfo{author}{Broderick, J.}, \bibinfo{author}{Broekema, P.C.}, \bibinfo{author}{Brouw, W.N.}, \bibinfo{author}{Brüggen, M.}, \bibinfo{author}{Butcher, H.R.}, \bibinfo{author}{van Cappellen, W.}, \bibinfo{author}{Ciardi, B.}, \bibinfo{author}{Coenen, T.}, \bibinfo{author}{Conway, J.}, \bibinfo{author}{Coolen, A.}, \bibinfo{author}{Corstanje, A.}, \bibinfo{author}{Damstra, S.}, \bibinfo{author}{Davies, O.}, \bibinfo{author}{Deller, A.T.}, \bibinfo{author}{Dettmar, R.J.}, \bibinfo{author}{van Diepen, G.}, \bibinfo{author}{Dijkstra, K.}, \bibinfo{author}{Donker, P.}, \bibinfo{author}{Doorduin, A.}, \bibinfo{author}{Dromer, J.}, \bibinfo{author}{Drost, M.}, \bibinfo{author}{van Duin, A.}, \bibinfo{author}{Eislöffel, J.}, \bibinfo{author}{van Enst, J.}, \bibinfo{author}{Ferrari, C.}, \bibinfo{author}{Frieswijk, W.}, \bibinfo{author}{Gankema, H.}, \bibinfo{author}{Garrett, M.A.}, \bibinfo{author}{de~Gasperin, F.}, \bibinfo{author}{Gerbers, M.}, \bibinfo{author}{de~Geus, E.}, \bibinfo{author}{Grie{\ss}meier, J.M.}, \bibinfo{author}{Grit, T.}, \bibinfo{author}{Gruppen, P.}, \bibinfo{author}{Hamaker, J.P.}, \bibinfo{author}{Hassall, T.}, \bibinfo{author}{Hoeft, M.}, \bibinfo{author}{Holties, H.A.}, \bibinfo{author}{Horneffer, A.}, \bibinfo{author}{van~der Horst, A.}, \bibinfo{author}{van Houwelingen, A.}, \bibinfo{author}{Huijgen, A.}, \bibinfo{author}{Iacobelli, M.}, \bibinfo{author}{Intema, H.}, \bibinfo{author}{Jackson, N.}, \bibinfo{author}{Jelic, V.}, \bibinfo{author}{de~Jong, A.}, \bibinfo{author}{Juette, E.}, \bibinfo{author}{Kant, D.}, \bibinfo{author}{Karastergiou, A.}, \bibinfo{author}{Koers, A.}, \bibinfo{author}{Kollen, H.}, \bibinfo{author}{Kondratiev, V.I.}, \bibinfo{author}{Kooistra, E.}, \bibinfo{author}{Koopman, Y.}, \bibinfo{author}{Koster, A.}, \bibinfo{author}{Kuniyoshi, M.}, \bibinfo{author}{Kramer, M.}, \bibinfo{author}{Kuper, G.}, \bibinfo{author}{Lambropoulos, P.}, \bibinfo{author}{Law, C.}, \bibinfo{author}{van Leeuwen, J.}, \bibinfo{author}{Lemaitre, J.}, \bibinfo{author}{Loose, M.}, \bibinfo{author}{Maat, P.}, \bibinfo{author}{Macario, G.}, \bibinfo{author}{Markoff, S.}, \bibinfo{author}{Masters, J.}, \bibinfo{author}{McFadden, R.A.}, \bibinfo{author}{McKay-Bukowski, D.}, \bibinfo{author}{Meijering, H.}, \bibinfo{author}{Meulman, H.}, \bibinfo{author}{Mevius, M.}, \bibinfo{author}{Middelberg, E.}, \bibinfo{author}{Millenaar, R.}, \bibinfo{author}{Miller-Jones, J.C.A.}, \bibinfo{author}{Mohan, R.N.}, \bibinfo{author}{Mol, J.D.}, \bibinfo{author}{Morawietz, J.}, \bibinfo{author}{Morganti, R.}, \bibinfo{author}{Mulcahy, D.D.}, \bibinfo{author}{Mulder, E.}, \bibinfo{author}{Munk, H.}, \bibinfo{author}{Nieuwenhuis, L.}, \bibinfo{author}{van Nieuwpoort, R.}, \bibinfo{author}{Noordam, J.E.}, \bibinfo{author}{Norden, M.}, \bibinfo{author}{Noutsos, A.}, \bibinfo{author}{Offringa, A.R.}, \bibinfo{author}{Olofsson, H.}, \bibinfo{author}{Omar, A.}, \bibinfo{author}{Orr{\'{u}}, E.}, \bibinfo{author}{Overeem, R.}, \bibinfo{author}{Paas, H.}, \bibinfo{author}{Pandey-Pommier, M.}, \bibinfo{author}{Pandey, V.N.}, \bibinfo{author}{Pizzo, R.}, \bibinfo{author}{Polatidis, A.}, \bibinfo{author}{Rafferty, D.}, \bibinfo{author}{Rawlings, S.}, \bibinfo{author}{Reich, W.}, \bibinfo{author}{de~Reijer, J.P.}, \bibinfo{author}{Reitsma, J.}, \bibinfo{author}{Renting, G.A.}, \bibinfo{author}{Riemers, P.}, \bibinfo{author}{Rol, E.}, \bibinfo{author}{Romein, J.W.}, \bibinfo{author}{Roosjen, J.}, \bibinfo{author}{Ruiter, M.}, \bibinfo{author}{Scaife, A.}, \bibinfo{author}{van~der Schaaf, K.}, \bibinfo{author}{Scheers, B.}, \bibinfo{author}{Schellart, P.}, \bibinfo{author}{Schoenmakers, A.}, \bibinfo{author}{Schoonderbeek, G.}, \bibinfo{author}{Serylak, M.}, \bibinfo{author}{Shulevski, A.}, \bibinfo{author}{Sluman, J.}, \bibinfo{author}{Smirnov, O.}, \bibinfo{author}{Sobey, C.}, \bibinfo{author}{Spreeuw, H.}, \bibinfo{author}{Steinmetz, M.}, \bibinfo{author}{Sterks, C.G.M.}, \bibinfo{author}{Stiepel, H.J.}, \bibinfo{author}{Stuurwold, K.}, \bibinfo{author}{Tagger, M.}, \bibinfo{author}{Tang, Y.}, \bibinfo{author}{Tasse, C.}, \bibinfo{author}{Thomas, I.}, \bibinfo{author}{Thoudam, S.}, \bibinfo{author}{Toribio, M.C.}, \bibinfo{author}{van~der Tol, B.}, \bibinfo{author}{Usov, O.}, \bibinfo{author}{van Veelen, M.}, \bibinfo{author}{van~der Veen, A.J.}, \bibinfo{author}{ter Veen, S.}, \bibinfo{author}{Verbiest, J.P.W.}, \bibinfo{author}{Vermeulen, R.}, \bibinfo{author}{Vermaas, N.}, \bibinfo{author}{Vocks, C.}, \bibinfo{author}{Vogt, C.}, \bibinfo{author}{de~Vos, M.}, \bibinfo{author}{van~der Wal, E.}, \bibinfo{author}{van Weeren, R.}, \bibinfo{author}{Weggemans, H.}, \bibinfo{author}{Weltevrede, P.}, \bibinfo{author}{White, S.}, \bibinfo{author}{Wijnholds, S.J.}, \bibinfo{author}{Wilhelmsson, T.}, \bibinfo{author}{Wucknitz, O.}, \bibinfo{author}{Yatawatta, S.}, \bibinfo{author}{Zarka, P.}, \bibinfo{author}{Zensus, A.}, \bibinfo{author}{van Zwieten, J.}, \bibinfo{year}{2013}.
\newblock \bibinfo{title}{{LOFAR}: The {LOw}-frequency {ARray}}.
\newblock \bibinfo{journal}{Astronomy and Astrophysics} \bibinfo{volume}{556}, \bibinfo{pages}{A2}.
\newblock \DOIprefix\doi{10.1051/0004-6361/201220873}.
\bibitem[{{Hallinan} et~al.(2019){Hallinan}, {Ravi}, {Weinreb}, {Kocz}, {Huang}, {Woody}, {Lamb}, {D'Addario}, {Catha}, {Law}, {Kulkarni}, {Phinney}, {Eastwood}, {Bouman}, {McLaughlin}, {Ransom}, {Siemens}, {Cordes}, {Lynch}, {Kaplan}, {Brazier}, {Bhatnagar}, {Myers}, {Walter} and {Gaensler}}]{dsa2000}
\bibinfo{author}{{Hallinan}, G.}, \bibinfo{author}{{Ravi}, V.}, \bibinfo{author}{{Weinreb}, S.}, \bibinfo{author}{{Kocz}, J.}, \bibinfo{author}{{Huang}, Y.}, \bibinfo{author}{{Woody}, D.P.}, \bibinfo{author}{{Lamb}, J.}, \bibinfo{author}{{D'Addario}, L.}, \bibinfo{author}{{Catha}, M.}, \bibinfo{author}{{Law}, C.}, \bibinfo{author}{{Kulkarni}, S.R.}, \bibinfo{author}{{Phinney}, E.S.}, \bibinfo{author}{{Eastwood}, M.W.}, \bibinfo{author}{{Bouman}, K.}, \bibinfo{author}{{McLaughlin}, M.}, \bibinfo{author}{{Ransom}, S.}, \bibinfo{author}{{Siemens}, X.}, \bibinfo{author}{{Cordes}, J.}, \bibinfo{author}{{Lynch}, R.}, \bibinfo{author}{{Kaplan}, D.}, \bibinfo{author}{{Brazier}, A.}, \bibinfo{author}{{Bhatnagar}, S.}, \bibinfo{author}{{Myers}, S.}, \bibinfo{author}{{Walter}, F.}, \bibinfo{author}{{Gaensler}, B.}, \bibinfo{year}{2019}.
\newblock \bibinfo{title}{{The DSA-2000 {\textemdash} A Radio Survey Camera}}, in: \bibinfo{booktitle}{Bulletin of the American Astronomical Society}, p. \bibinfo{pages}{255}.
\newblock \DOIprefix\doi{10.48550/arXiv.1907.07648}, \href{http://arxiv.org/abs/1907.07648}{{\tt arXiv:1907.07648}}.
\bibitem[{Hamman and Hoyer(2017)}]{xarray2017}
\bibinfo{author}{Hamman, J.}, \bibinfo{author}{Hoyer, S.}, \bibinfo{year}{2017}.
\newblock \bibinfo{title}{Xarray: N-d labeled arrays and datasets in python}.
\newblock \bibinfo{journal}{Journal of Open Research Software} .
\bibitem[{Harris et~al.(2020)Harris, Millman, van~der Walt, Gommers, Virtanen, Cournapeau, Wieser, Taylor, Berg, Smith, Kern, Picus, Hoyer, van Kerkwijk, Brett, Haldane, del R{'{\i}}o, Wiebe, Peterson, G{'{e}}rard-Marchant, Sheppard, Reddy, Weckesser, Abbasi, Gohlke and Oliphant}]{Harris2020}
\bibinfo{author}{Harris, C.R.}, \bibinfo{author}{Millman, K.J.}, \bibinfo{author}{van~der Walt, S.J.}, \bibinfo{author}{Gommers, R.}, \bibinfo{author}{Virtanen, P.}, \bibinfo{author}{Cournapeau, D.}, \bibinfo{author}{Wieser, E.}, \bibinfo{author}{Taylor, J.}, \bibinfo{author}{Berg, S.}, \bibinfo{author}{Smith, N.J.}, \bibinfo{author}{Kern, R.}, \bibinfo{author}{Picus, M.}, \bibinfo{author}{Hoyer, S.}, \bibinfo{author}{van Kerkwijk, M.H.}, \bibinfo{author}{Brett, M.}, \bibinfo{author}{Haldane, A.}, \bibinfo{author}{del R{'{\i}}o, J.F.}, \bibinfo{author}{Wiebe, M.}, \bibinfo{author}{Peterson, P.}, \bibinfo{author}{G{'{e}}rard-Marchant, P.}, \bibinfo{author}{Sheppard, K.}, \bibinfo{author}{Reddy, T.}, \bibinfo{author}{Weckesser, W.}, \bibinfo{author}{Abbasi, H.}, \bibinfo{author}{Gohlke, C.}, \bibinfo{author}{Oliphant, T.E.}, \bibinfo{year}{2020}.
\newblock \bibinfo{title}{Array programming with {NumPy}}.
\newblock \bibinfo{journal}{Nature} \bibinfo{volume}{585}, \bibinfo{pages}{357--362}.
\newblock \DOIprefix\doi{10.1038/s41586-020-2649-2}.
\bibitem[{Hewitt et~al.(1973)Hewitt, Bishop and Steiger}]{Hewitt1973}
\bibinfo{author}{Hewitt, C.}, \bibinfo{author}{Bishop, P.}, \bibinfo{author}{Steiger, R.}, \bibinfo{year}{1973}.
\newblock \bibinfo{title}{A universal modular actor formalism for artificial intelligence}, in: \bibinfo{booktitle}{Proceedings of the 3rd International Joint Conference on Artificial Intelligence}, \bibinfo{publisher}{Morgan Kaufmann Publishers Inc.}, \bibinfo{address}{San Francisco, CA, USA}. p. \bibinfo{pages}{235–245}.
\bibitem[{Hotan et~al.(2021)Hotan, Bunton, Chippendale, Whiting, Tuthill, Moss, McConnell, Amy, Huynh, Allison, Anderson, Bannister, Bastholm, Beresford, Bock, Bolton, Chapman, Chow, Collier, Cooray, Cornwell, Diamond, Edwards, Feain, Franzen, George, Gupta, Hampson, Harvey-Smith, Hayman, Heywood, Jacka, Jackson, Jackson, Jeganathan, Johnston, Kesteven, Kleiner, Koribalski, Lee-Waddell, Lenc, Lensson, Mackay, Mahony, McClure-Griffiths, McConigley, Mirtschin, Ng, Norris, Pearce, Phillips, Pilawa, Raja, Reynolds, Roberts, Roxby, Sadler, Shields, Schinckel, Serra, Shaw, Sweetnam, Troup, Tzioumis, Voronkov and Westmeier}]{Hotan_2021}
\bibinfo{author}{Hotan, A.W.}, \bibinfo{author}{Bunton, J.D.}, \bibinfo{author}{Chippendale, A.P.}, \bibinfo{author}{Whiting, M.}, \bibinfo{author}{Tuthill, J.}, \bibinfo{author}{Moss, V.A.}, \bibinfo{author}{McConnell, D.}, \bibinfo{author}{Amy, S.W.}, \bibinfo{author}{Huynh, M.T.}, \bibinfo{author}{Allison, J.R.}, \bibinfo{author}{Anderson, C.S.}, \bibinfo{author}{Bannister, K.W.}, \bibinfo{author}{Bastholm, E.}, \bibinfo{author}{Beresford, R.}, \bibinfo{author}{Bock, D.C.J.}, \bibinfo{author}{Bolton, R.}, \bibinfo{author}{Chapman, J.M.}, \bibinfo{author}{Chow, K.}, \bibinfo{author}{Collier, J.D.}, \bibinfo{author}{Cooray, F.R.}, \bibinfo{author}{Cornwell, T.J.}, \bibinfo{author}{Diamond, P.J.}, \bibinfo{author}{Edwards, P.G.}, \bibinfo{author}{Feain, I.J.}, \bibinfo{author}{Franzen, T.M.O.}, \bibinfo{author}{George, D.}, \bibinfo{author}{Gupta, N.}, \bibinfo{author}{Hampson, G.A.}, \bibinfo{author}{Harvey-Smith, L.}, \bibinfo{author}{Hayman, D.B.}, \bibinfo{author}{Heywood, I.}, \bibinfo{author}{Jacka, C.}, \bibinfo{author}{Jackson, C.A.}, \bibinfo{author}{Jackson, S.}, \bibinfo{author}{Jeganathan, K.}, \bibinfo{author}{Johnston, S.}, \bibinfo{author}{Kesteven, M.}, \bibinfo{author}{Kleiner, D.}, \bibinfo{author}{Koribalski, B.S.}, \bibinfo{author}{Lee-Waddell, K.}, \bibinfo{author}{Lenc, E.}, \bibinfo{author}{Lensson, E.S.}, \bibinfo{author}{Mackay, S.}, \bibinfo{author}{Mahony, E.K.}, \bibinfo{author}{McClure-Griffiths, N.M.}, \bibinfo{author}{McConigley, R.}, \bibinfo{author}{Mirtschin, P.}, \bibinfo{author}{Ng, A.K.}, \bibinfo{author}{Norris, R.P.}, \bibinfo{author}{Pearce, S.E.}, \bibinfo{author}{Phillips, C.}, \bibinfo{author}{Pilawa, M.A.}, \bibinfo{author}{Raja, W.}, \bibinfo{author}{Reynolds, J.E.}, \bibinfo{author}{Roberts, P.}, \bibinfo{author}{Roxby, D.N.}, \bibinfo{author}{Sadler, E.M.}, \bibinfo{author}{Shields, M.}, \bibinfo{author}{Schinckel, A.E.T.}, \bibinfo{author}{Serra, P.}, \bibinfo{author}{Shaw, R.D.}, \bibinfo{author}{Sweetnam, T.}, \bibinfo{author}{Troup, E.R.}, \bibinfo{author}{Tzioumis, A.}, \bibinfo{author}{Voronkov, M.A.}, \bibinfo{author}{Westmeier, T.}, \bibinfo{year}{2021}.
\newblock \bibinfo{title}{Australian square kilometre array pathfinder: I. system description}.
\newblock \bibinfo{journal}{Publications of the Astronomical Society of Australia} \bibinfo{volume}{38}.
\newblock \DOIprefix\doi{10.1017/pasa.2021.1}.
\bibitem[{Hugo(2024)}]{Hugo2024}
\bibinfo{author}{Hugo, B.}, \bibinfo{year}{2024}.
\newblock \bibinfo{title}{A Journey from Commissioning to Science with MeerKAT}.
\newblock Ph.D. thesis. Rhodes University.
\bibitem[{Hugo et~al.(2022)Hugo, Perkins, Merry, Mauch and Smirnov}]{Hugo2022}
\bibinfo{author}{Hugo, B.V.}, \bibinfo{author}{Perkins, S.}, \bibinfo{author}{Merry, B.}, \bibinfo{author}{Mauch, T.}, \bibinfo{author}{Smirnov, O.M.}, \bibinfo{year}{2022}.
\newblock \bibinfo{title}{Tricolour: an optimized sumthreshold flagger for meerkat}, in: \bibinfo{editor}{{Ruiz}, J.E.}, \bibinfo{editor}{{Pierfedereci}, F.}, \bibinfo{editor}{{Teuben}, P.} (Eds.), \bibinfo{booktitle}{Astronomical Society of the Pacific Conference Series}.
\bibitem[{Hunter(2007)}]{matplotlib2007}
\bibinfo{author}{Hunter, J.D.}, \bibinfo{year}{2007}.
\newblock \bibinfo{title}{Matplotlib: A 2d graphics environment}.
\newblock \bibinfo{journal}{Computing in Science \& Engineering} \bibinfo{volume}{9}, \bibinfo{pages}{90--95}.
\newblock \DOIprefix\doi{10.1109/MCSE.2007.55}.
\bibitem[{Jakob et~al.(2017)Jakob, Rhinelander and Moldovan}]{pybind11}
\bibinfo{author}{Jakob, W.}, \bibinfo{author}{Rhinelander, J.}, \bibinfo{author}{Moldovan, D.}, \bibinfo{year}{2017}.
\newblock \bibinfo{title}{{pybind11 -- Seamless operability between C++11 and Python}}.
\newblock \bibinfo{note}{Https://github.com/pybind/pybind11}.
\bibitem[{{Jonas} and {MeerKAT Team}(2016)}]{Jonas2016}
\bibinfo{author}{{Jonas}, J.}, \bibinfo{author}{{MeerKAT Team}}, \bibinfo{year}{2016}.
\newblock \bibinfo{title}{{The MeerKAT Radio Telescope}}, in: \bibinfo{booktitle}{MeerKAT Science: On the Pathway to the SKA}, p.~\bibinfo{pages}{1}.
\newblock \DOIprefix\doi{10.22323/1.277.0001}.
\bibitem[{{J{\'o}zsa} et~al.(2022){J{\'o}zsa}, {Andati}, {de Blok}, {Hugo}, {Kleiner}, {Kamphuis}, {Moln{\'a}r}, {Makhathini}, {Maccagni}, {Perkins}, {Ramaila}, {Ramatsoku}, {Serra}, {Smirnov}, {Thorat} and {White}}]{caracal}
\bibinfo{author}{{J{\'o}zsa}, G.I.G.}, \bibinfo{author}{{Andati}, L.A.L.}, \bibinfo{author}{{de Blok}, W.J.G.}, \bibinfo{author}{{Hugo}, B.V.}, \bibinfo{author}{{Kleiner}, D.}, \bibinfo{author}{{Kamphuis}, P.}, \bibinfo{author}{{Moln{\'a}r}, D.C.}, \bibinfo{author}{{Makhathini}, S.}, \bibinfo{author}{{Maccagni}, F.M.}, \bibinfo{author}{{Perkins}, S.J.}, \bibinfo{author}{{Ramaila}, A.}, \bibinfo{author}{{Ramatsoku}, M.}, \bibinfo{author}{{Serra}, P.}, \bibinfo{author}{{Smirnov}, O.M.}, \bibinfo{author}{{Thorat}, K.}, \bibinfo{author}{{White}, S.V.}, \bibinfo{year}{2022}.
\newblock \bibinfo{title}{{CARACal - The Containerized Automated Radio Astronomy Calibration Pipeline}}, in: \bibinfo{editor}{{Ruiz}, J.E.}, \bibinfo{editor}{{Pierfedereci}, F.}, \bibinfo{editor}{{Teuben}, P.} (Eds.), \bibinfo{booktitle}{Astronomical Society of the Pacific Conference Series}, p. \bibinfo{pages}{447}.
\bibitem[{{Kemball} and {Wieringa}(2000)}]{MeasurementSet2}
\bibinfo{author}{{Kemball}, A.J.}, \bibinfo{author}{{Wieringa}, M.H.}, \bibinfo{year}{2000}.
\newblock \bibinfo{title}{{Measurement Set definition version 2.0}}.
\newblock \bibinfo{howpublished}{NRAO report, January 21, 2000, 52 pages}.
\bibitem[{Kenyon et~al.(2024)Kenyon, Perkins, Bester, Smirnov, Russeeawon and Hugo}]{africanus2}
\bibinfo{author}{Kenyon, J.S.}, \bibinfo{author}{Perkins, S.J.}, \bibinfo{author}{Bester, H.L.}, \bibinfo{author}{Smirnov, O.M.}, \bibinfo{author}{Russeeawon, C.}, \bibinfo{author}{Hugo, B.V.}, \bibinfo{year}{2024}.
\newblock \bibinfo{title}{{Africanus II. QuartiCal: calibrating radio interferometer data at scale using Numba and Dask}}.
\newblock \bibinfo{journal}{Astronomy and Computing} \bibinfo{volume}{submitted}.
\newblock \href{http://arxiv.org/abs/2412.10072}{{\tt arXiv:2412.10072}}.
\bibitem[{{Kenyon} et~al.(2018){Kenyon}, {Smirnov}, {Grobler} and {Perkins}}]{kenyon2018cubical}
\bibinfo{author}{{Kenyon}, J.S.}, \bibinfo{author}{{Smirnov}, O.M.}, \bibinfo{author}{{Grobler}, T.L.}, \bibinfo{author}{{Perkins}, S.J.}, \bibinfo{year}{2018}.
\newblock \bibinfo{title}{{CUBICAL - fast radio interferometric calibration suite exploiting complex optimization}}.
\newblock \bibinfo{journal}{Monthly Notices of the Royal Astronomical Society} \bibinfo{volume}{478}, \bibinfo{pages}{2399--2415}.
\newblock \DOIprefix\doi{10.1093/mnras/sty1221}, \href{http://arxiv.org/abs/1805.03410}{{\tt arXiv:1805.03410}}.
\bibitem[{{Kettenis} et~al.(2006){Kettenis}, {van Langevelde}, {Reynolds} and {Cotton}}]{Kettenis2006}
\bibinfo{author}{{Kettenis}, M.}, \bibinfo{author}{{van Langevelde}, H.J.}, \bibinfo{author}{{Reynolds}, C.}, \bibinfo{author}{{Cotton}, B.}, \bibinfo{year}{2006}.
\newblock \bibinfo{title}{{ParselTongue: AIPS Talking Python}}, in: \bibinfo{editor}{{Gabriel}, C.}, \bibinfo{editor}{{Arviset}, C.}, \bibinfo{editor}{{Ponz}, D.}, \bibinfo{editor}{{Enrique}, S.} (Eds.), \bibinfo{booktitle}{Astronomical Data Analysis Software and Systems XV}, p. \bibinfo{pages}{497}.
\bibitem[{Kimball and Caserta(2004)}]{Kimball2004}
\bibinfo{author}{Kimball, R.}, \bibinfo{author}{Caserta, J.}, \bibinfo{year}{2004}.
\newblock \bibinfo{title}{The Data Warehouse ETL Toolkit: Practical Techniques for Extracting, Cleaning, Conforming, and Delivering Data}.
\newblock \bibinfo{publisher}{Wiley}, \bibinfo{address}{Indianapolis, IN}.
\bibitem[{Kwok and Ahmad(1999)}]{Kwok1999}
\bibinfo{author}{Kwok, Y.K.}, \bibinfo{author}{Ahmad, I.}, \bibinfo{year}{1999}.
\newblock \bibinfo{title}{Static scheduling algorithms for allocating directed task graphs to multiprocessors}.
\newblock \bibinfo{journal}{ACM Comput. Surv.} \bibinfo{volume}{31}, \bibinfo{pages}{406–471}.
\newblock \URLprefix \url{https://doi.org/10.1145/344588.344618}, \DOIprefix\doi{10.1145/344588.344618}.
\bibitem[{Lam et~al.(2015)Lam, Pitrou and Seibert}]{Lam2015}
\bibinfo{author}{Lam, S.K.}, \bibinfo{author}{Pitrou, A.}, \bibinfo{author}{Seibert, S.}, \bibinfo{year}{2015}.
\newblock \bibinfo{title}{Numba: A llvm-based python jit compiler}, in: \bibinfo{booktitle}{Proceedings of the Second Workshop on the LLVM Compiler Infrastructure in HPC}, \bibinfo{publisher}{Association for Computing Machinery}, \bibinfo{address}{New York, NY, USA}.
\newblock \DOIprefix\doi{10.1145/2833157.2833162}.
\bibitem[{Makhathini(2018)}]{makhathini2018}
\bibinfo{author}{Makhathini, S.}, \bibinfo{year}{2018}.
\newblock \bibinfo{title}{Advanced radio interferometric simulation and data reduction techniques}.
\newblock Ph.D. thesis. Rhodes University. \bibinfo{address}{Drosty Rd, Grahamstown, 6139, Eastern Cape, South Africa}.
\newblock \bibinfo{note}{Available via \url{http://hdl.handle.net/10962/57348}}.
\bibitem[{{W}es {M}c{K}inney(2010)}]{pandas2010}
\bibinfo{author}{{W}es {M}c{K}inney}, \bibinfo{year}{2010}.
\newblock \bibinfo{title}{{D}ata {S}tructures for {S}tatistical {C}omputing in {P}ython}, in: \bibinfo{editor}{{S}t\'efan van~der {W}alt}, \bibinfo{editor}{{J}arrod {M}illman} (Eds.), \bibinfo{booktitle}{{P}roceedings of the 9th {P}ython in {S}cience {C}onference}, pp. \bibinfo{pages}{56 -- 61}.
\newblock \DOIprefix\doi{10.25080/Majora-92bf1922-00a}.
\bibitem[{{McMullin} et~al.(2007){McMullin}, {Waters}, {Schiebel}, {Young} and {Golap}}]{McMullin2007}
\bibinfo{author}{{McMullin}, J.P.}, \bibinfo{author}{{Waters}, B.}, \bibinfo{author}{{Schiebel}, D.}, \bibinfo{author}{{Young}, W.}, \bibinfo{author}{{Golap}, K.}, \bibinfo{year}{2007}.
\newblock \bibinfo{title}{{CASA Architecture and Applications}}, in: \bibinfo{editor}{{Shaw}, R.A.}, \bibinfo{editor}{{Hill}, F.}, \bibinfo{editor}{{Bell}, D.J.} (Eds.), \bibinfo{booktitle}{Astronomical Data Analysis Software and Systems XVI}, p. \bibinfo{pages}{127}.
\bibitem[{{Message Passing Interface Forum}(2023)}]{mpi41}
\bibinfo{author}{{Message Passing Interface Forum}}, \bibinfo{year}{2023}.
\newblock \bibinfo{title}{{MPI}: A Message-Passing Interface Standard Version 4.1}.
\newblock \URLprefix \url{https://www.mpi-forum.org/docs/mpi-4.1/mpi41-report.pdf}.
\bibitem[{Molenaar(2021)}]{Molenaar2021}
\bibinfo{author}{Molenaar, G.J.}, \bibinfo{year}{2021}.
\newblock \bibinfo{title}{Design patterns and software techniques for large-scale, open and reproducible data reduction}.
\newblock Ph.D. thesis. Rhodes University.
\bibitem[{{Noordam} and {Smirnov}(2012)}]{Noordam2012}
\bibinfo{author}{{Noordam}, J.E.}, \bibinfo{author}{{Smirnov}, O.M.}, \bibinfo{year}{2012}.
\newblock \bibinfo{title}{{MeqTrees: Software package for implementing Measurement Equations}}.
\newblock \href{http://arxiv.org/abs/1209.010}{{\tt arXiv:1209.010}}.
\bibitem[{Norman(2002)}]{Norman2002}
\bibinfo{author}{Norman, D.A.}, \bibinfo{year}{2002}.
\newblock \bibinfo{title}{The Design of Everyday Things}.
\newblock \bibinfo{publisher}{Basic Books, Inc.}, \bibinfo{address}{USA}.
\bibitem[{{Offringa}(2010)}]{AOFlagger2010}
\bibinfo{author}{{Offringa}, A.R.}, \bibinfo{year}{2010}.
\newblock \bibinfo{title}{Aoflagger: Rfi software}.
\newblock \bibinfo{howpublished}{Astrophysics Source Code Library, record ascl:1010.017}.
\newblock \href{http://arxiv.org/abs/1010.017}{{\tt arXiv:1010.017}}.
\bibitem[{{Offringa}(2012)}]{offringa2012}
\bibinfo{author}{{Offringa}, A.R.}, \bibinfo{year}{2012}.
\newblock \bibinfo{title}{The SumThreshold Method: Technical Details}.
\newblock \bibinfo{type}{Technical Report}. Rijksuniversiteit Groningen.
\bibitem[{{O'Mullane} et~al.(2023){O'Mullane}, {Guy}, {Dubois-Felsmann}, {Economou}, {AlSayyad}, {Graham} and {Slater}}]{rubin2023}
\bibinfo{author}{{O'Mullane}, W.}, \bibinfo{author}{{Guy}, L.}, \bibinfo{author}{{Dubois-Felsmann}, G.}, \bibinfo{author}{{Economou}, F.}, \bibinfo{author}{{AlSayyad}, Y.}, \bibinfo{author}{{Graham}, M.}, \bibinfo{author}{{Slater}, C.}, \bibinfo{year}{2023}.
\newblock \bibinfo{title}{{Vera C. Rubin Observatory: Open Science to the core}}, in: \bibinfo{booktitle}{American Astronomical Society Meeting Abstracts}, p. \bibinfo{pages}{116.04}.
\bibitem[{{OpenXLA Contributors}(2024)}]{OpenXLA}
\bibinfo{author}{{OpenXLA Contributors}}, \bibinfo{year}{2024}.
\newblock \bibinfo{title}{Openxla: A unified compiler ecosystem for ml workloads}.
\newblock \URLprefix \url{https://openxla.org}.
\bibitem[{{Perkins} et~al.(2015){Perkins}, {Marais}, {Zwart}, {Natarajan}, {Tasse} and {Smirnov}}]{Perkins2015}
\bibinfo{author}{{Perkins}, S.J.}, \bibinfo{author}{{Marais}, P.C.}, \bibinfo{author}{{Zwart}, J.T.L.}, \bibinfo{author}{{Natarajan}, I.}, \bibinfo{author}{{Tasse}, C.}, \bibinfo{author}{{Smirnov}, O.}, \bibinfo{year}{2015}.
\newblock \bibinfo{title}{{Montblanc$^{1}$: GPU accelerated radio interferometer measurement equations in support of Bayesian inference for radio observations}}.
\newblock \bibinfo{journal}{Astronomy and Computing} \bibinfo{volume}{12}, \bibinfo{pages}{73--85}.
\newblock \DOIprefix\doi{10.1016/j.ascom.2015.06.003}, \href{http://arxiv.org/abs/1501.07719}{{\tt arXiv:1501.07719}}.
\bibitem[{{Ramatsoku, M.} et~al.(2020){Ramatsoku, M.}, {Murgia, M.}, {Vacca, V.}, {Serra, P.}, {Makhathini, S.}, {Govoni, F.}, {Smirnov, O.}, {Andati, L. A. L.}, {de Blok, E.}, {J\'ozsa, G. I. G.}, {Kamphuis, P.}, {Kleiner, D.}, {Maccagni, F. M.}, {Moln\'ar, D. Cs.}, {Ramaila, A. J. T.}, {Thorat, K.} and {White, S. V.}}]{Ramatsoku2020}
\bibinfo{author}{{Ramatsoku, M.}}, \bibinfo{author}{{Murgia, M.}}, \bibinfo{author}{{Vacca, V.}}, \bibinfo{author}{{Serra, P.}}, \bibinfo{author}{{Makhathini, S.}}, \bibinfo{author}{{Govoni, F.}}, \bibinfo{author}{{Smirnov, O.}}, \bibinfo{author}{{Andati, L. A. L.}}, \bibinfo{author}{{de Blok, E.}}, \bibinfo{author}{{J\'ozsa, G. I. G.}}, \bibinfo{author}{{Kamphuis, P.}}, \bibinfo{author}{{Kleiner, D.}}, \bibinfo{author}{{Maccagni, F. M.}}, \bibinfo{author}{{Moln\'ar, D. Cs.}}, \bibinfo{author}{{Ramaila, A. J. T.}}, \bibinfo{author}{{Thorat, K.}}, \bibinfo{author}{{White, S. V.}}, \bibinfo{year}{2020}.
\newblock \bibinfo{title}{Collimated synchrotron threads linking the radio lobes of eso 137-006}.
\newblock \bibinfo{journal}{A\&A} \bibinfo{volume}{636}, \bibinfo{pages}{L1}.
\newblock \URLprefix \url{https://doi.org/10.1051/0004-6361/202037800}, \DOIprefix\doi{10.1051/0004-6361/202037800}.
\bibitem[{Rodrigues and Rodriguez~Bezos(1969)}]{Rodrigues1969}
\bibinfo{author}{Rodrigues, J.E.}, \bibinfo{author}{Rodriguez~Bezos, J.E.}, \bibinfo{year}{1969}.
\newblock \bibinfo{title}{A GRAPH MODEL FOR PARALLEL COMPUTATIONS}.
\newblock \bibinfo{type}{Technical Report}. Massachusetts Institute of Technology. \bibinfo{address}{USA}.
\bibitem[{Sabater et~al.(2017)Sabater, Sánchez-Expósito, Best, Garrido, Verdes-Montenegro and Lezzi}]{cloudlofar2017}
\bibinfo{author}{Sabater, J.}, \bibinfo{author}{Sánchez-Expósito, S.}, \bibinfo{author}{Best, P.}, \bibinfo{author}{Garrido, J.}, \bibinfo{author}{Verdes-Montenegro, L.}, \bibinfo{author}{Lezzi, D.}, \bibinfo{year}{2017}.
\newblock \bibinfo{title}{Calibration of lofar data on the cloud}.
\newblock \bibinfo{journal}{Astronomy and Computing} \bibinfo{volume}{19}, \bibinfo{pages}{75--89}.
\newblock \DOIprefix\doi{https://doi.org/10.1016/j.ascom.2017.04.001}.
\bibitem[{Schilizzi et~al.(2008)Schilizzi, Dewdney and Lazio}]{schilizzi2008square}
\bibinfo{author}{Schilizzi, R.T.}, \bibinfo{author}{Dewdney, P.E.}, \bibinfo{author}{Lazio, T.J.W.}, \bibinfo{year}{2008}.
\newblock \bibinfo{title}{The square kilometre array}, in: \bibinfo{booktitle}{SPIE Astronomical Telescopes+ Instrumentation}, \bibinfo{organization}{International Society for Optics and Photonics}. pp. \bibinfo{pages}{70121I--70121I}.
\bibitem[{{Schwardt} et~al.(2023){Schwardt}, {Merry}, {Perkins}, {Richter}, {Mauch} and {Ratcliffe}}]{Schwardt2023}
\bibinfo{author}{{Schwardt}, L.}, \bibinfo{author}{{Merry}, B.}, \bibinfo{author}{{Perkins}, S.}, \bibinfo{author}{{Richter}, L.}, \bibinfo{author}{{Mauch}, T.}, \bibinfo{author}{{Ratcliffe}, S.}, \bibinfo{year}{2023}.
\newblock \bibinfo{title}{{katdal: MeerKAT Data Access Library}}.
\newblock \bibinfo{howpublished}{Astrophysics Source Code Library, record ascl:2305.004}.
\bibitem[{Sekhar and Athreya(2018)}]{Sekhar2018}
\bibinfo{author}{Sekhar, S.}, \bibinfo{author}{Athreya, R.}, \bibinfo{year}{2018}.
\newblock \bibinfo{title}{Two procedures to flag radio frequency interference in the uv plane}.
\newblock \bibinfo{journal}{The Astronomical Journal} \bibinfo{volume}{156}, \bibinfo{pages}{9}.
\newblock \URLprefix \url{https://dx.doi.org/10.3847/1538-3881/aac16e}, \DOIprefix\doi{10.3847/1538-3881/aac16e}.
\bibitem[{{Smirnov}(2011)}]{Smirnov2011}
\bibinfo{author}{{Smirnov}, O.M.}, \bibinfo{year}{2011}.
\newblock \bibinfo{title}{{Revisiting the radio interferometer measurement equation - I. A full-sky Jones formalism}}.
\newblock \bibinfo{journal}{Astronomy and Astrophysics} \bibinfo{volume}{527}, \bibinfo{pages}{A106}.
\newblock \DOIprefix\doi{10.1051/0004-6361/201016082}.
\bibitem[{{Smirnov} et~al.(2022){Smirnov}, {Heywood}, {Perkins} and {van Rooyen}}]{shadems2022}
\bibinfo{author}{{Smirnov}, O.M.}, \bibinfo{author}{{Heywood}, I.}, \bibinfo{author}{{Perkins}, S.J.}, \bibinfo{author}{{van Rooyen}, R.}, \bibinfo{year}{2022}.
\newblock \bibinfo{title}{{ShadeMS: Rapid Plotting of Big Radio Interferometry Data}}, in: \bibinfo{editor}{{Ruiz}, J.E.}, \bibinfo{editor}{{Pierfedereci}, F.}, \bibinfo{editor}{{Teuben}, P.} (Eds.), \bibinfo{booktitle}{Astronomical Society of the Pacific Conference Series}, p. \bibinfo{pages}{385}.
\bibitem[{Smirnov et~al.(2024)Smirnov, Makhathini, Kenyon, Bester, Perkins, Ramaila and Hugo}]{africanus4}
\bibinfo{author}{Smirnov, O.M.}, \bibinfo{author}{Makhathini, S.}, \bibinfo{author}{Kenyon, J.S.}, \bibinfo{author}{Bester, H.L.}, \bibinfo{author}{Perkins, S.J.}, \bibinfo{author}{Ramaila, A.J.T.}, \bibinfo{author}{Hugo, B.V.}, \bibinfo{year}{2024}.
\newblock \bibinfo{title}{{Africanus IV. The Stimela2 framework: scalable and reproducible workflows, from local to cloud compute}}.
\newblock \bibinfo{journal}{Astronomy and Computing} \bibinfo{volume}{submitted}.
\newblock \href{http://arxiv.org/abs/2412.10080}{{\tt arXiv:2412.10080}}.
\bibitem[{Tasse et~al.(2018)Tasse, Hugo, Mirmont, Smirnov, Atemkeng, Bester, Hardcastle, Lakhoo, Perkins and Shimwell}]{ddfacet2018}
\bibinfo{author}{Tasse, C.}, \bibinfo{author}{Hugo, B.}, \bibinfo{author}{Mirmont, M.}, \bibinfo{author}{Smirnov, O.}, \bibinfo{author}{Atemkeng, M.}, \bibinfo{author}{Bester, L.}, \bibinfo{author}{Hardcastle, M.J.}, \bibinfo{author}{Lakhoo, R.}, \bibinfo{author}{Perkins, S.}, \bibinfo{author}{Shimwell, T.}, \bibinfo{year}{2018}.
\newblock \bibinfo{title}{Faceting for direction-dependent spectral deconvolution}.
\newblock \bibinfo{journal}{Astronomy and Astrophysics} \bibinfo{volume}{611}, \bibinfo{pages}{A87}.
\newblock \DOIprefix\doi{10.1051/0004-6361/201731474}.
\bibitem[{Tilmes(2011)}]{Tilmes2011}
\bibinfo{author}{Tilmes, C.}, \bibinfo{year}{2011}.
\newblock \bibinfo{title}{{Data Formats: Using Self-Describing Data Formats}}.
\newblock \bibinfo{type}{Technical Report}. NASA.
\bibitem[{Toomey et~al.(2017)Toomey, Benn, Chapman, Dai, Dempsey, Hobbs, Russell, Wang, Wang and Zic}]{cloudpulsar2017}
\bibinfo{author}{Toomey, L.}, \bibinfo{author}{Benn, D.}, \bibinfo{author}{Chapman, J.}, \bibinfo{author}{Dai, S.}, \bibinfo{author}{Dempsey, J.}, \bibinfo{author}{Hobbs, G.}, \bibinfo{author}{Russell, C.}, \bibinfo{author}{Wang, C.}, \bibinfo{author}{Wang, J.}, \bibinfo{author}{Zic, J.}, \bibinfo{year}{2017}.
\newblock \bibinfo{title}{Processing public pulsar astronomy data in the Amazon Cloud}.
\newblock \bibinfo{type}{Technical Report}. CSIRO.
\bibitem[{Ullman(1975)}]{Ullman1975}
\bibinfo{author}{Ullman, J.D.}, \bibinfo{year}{1975}.
\newblock \bibinfo{title}{Np-complete scheduling problems}.
\newblock \bibinfo{journal}{J. Comput. Syst. Sci.} \bibinfo{volume}{10}, \bibinfo{pages}{384–393}.
\newblock \URLprefix \url{https://doi.org/10.1016/S0022-0000(75)80008-0}, \DOIprefix\doi{10.1016/S0022-0000(75)80008-0}.
\bibitem[{Virtanen et~al.(2020)}]{Virtanen2020}
\bibinfo{author}{Virtanen, P.}, et~al., \bibinfo{year}{2020}.
\newblock \bibinfo{title}{{{SciPy} 1.0: Fundamental Algorithms for Scientific Computing in Python}}.
\newblock \bibinfo{journal}{Nature Methods} \bibinfo{volume}{17}, \bibinfo{pages}{261--272}.
\newblock \DOIprefix\doi{10.1038/s41592-019-0686-2}.
\bibitem[{{Wells} et~al.(1981){Wells}, {Greisen} and {Harten}}]{Wells1981}
\bibinfo{author}{{Wells}, D.C.}, \bibinfo{author}{{Greisen}, E.W.}, \bibinfo{author}{{Harten}, R.H.}, \bibinfo{year}{1981}.
\newblock \bibinfo{title}{{FITS - a Flexible Image Transport System}}.
\newblock \bibinfo{journal}{A\&A} \bibinfo{volume}{44}, \bibinfo{pages}{363}.

\end{thebibliography}







\end{document}